\documentclass{aa}  

\usepackage{graphicx}
\usepackage{txfonts}
\usepackage{booktabs}
\usepackage{pdflscape}
\usepackage{placeins}
\usepackage[outercaption]{sidecap}
\usepackage[colorlinks=true,
            linkcolor=blue,
            urlcolor=blue,
            citecolor=blue]{hyperref}
\newcommand{\Teff}{\ensuremath{T_\mathrm{eff}}}
\newcommand{\logg}{\ensuremath{\log g}}
\newcommand{\vrad}{\ensuremath{v_\mathrm{rad}}}
\newcommand{\vsini}{\ensuremath{v_\mathrm{rot}\sin i}}
\newcommand{\lognHE}{\ensuremath{\log (n_\mathrm{He}/n_\mathrm{H})}}

\newcommand{\teff}{$T_\mathrm{eff}$}
\newcommand{\neon}{{$^{22}\mathrm{Ne}$}}

\begin{document}

   \title{Stars on the ascending helium giant branch  }

   \subtitle{II. A metal-rich luminous helium sdO gone astray}

   \author{M.~Pritzkuleit
          \inst{1}
          \and
          M.~Dorsch
          \inst{1}
          \and 
          M.~Miller Bertolami
          \inst{2,3}
          \and
          S.~Geier
          \inst{1}
          \and
          C.~W.~Bradshaw
          \inst{4}
          \and
          S.~Nepal
          \inst{5}
          \and
          N.~Reindl
          \inst{6}
          \and
          H.~Dawson
          \inst{1}
          }
          %Authorlist not final

   \institute{Institut für Physik und Astronomie, Universität Potsdam,
              Karl-Liebknecht-Str. 24/25, D-14476 Potsdam, Germany\\
              \email{maxpri@astro.physik.uni-potsdam.de}
      \and
            Instituto de Astrofísica de La Plata, Consejo Nacional de Investigaciones Científicas y Técnicas Avenida Centenario (Paseo del Bosque) S/N, B1900FWA La Plata Argentina
      \and  
            Facultad de Ciencias Astronómicas y Geofísicas, Universidad Nacional de La Plata Avenida Centenario (Paseo del Bosque) S/N, B1900FWA La Plata Argentina  
      \and 
             Hamburger Sternwarte, University of Hamburg, Gojenbergsweg 112, D-21029 Hamburg, Germany
      \and 
             Sydney Institute for Astronomy, School of Physics, University of Sydney, NSW 2006, Australia
      \and 
             Zentrum für Astronomie der Universität Heidelberg, Landessternwarte, Königstuhl 12, D-69117 Heidelberg, Germany       
             }
   \date{}

% \abstract{}{}{}{}{} 
% 5 {} token are mandatory
 
  \abstract
   {Hot subdwarf stars with masses above $\sim0.7~M_\odot$ expand after core helium exhaustion and ascend the helium giant branch before entering the white dwarf (WD) cooling track. In the first paper of this series, we identified the luminous helium-rich hot subdwarf HE~1518-0948 (HE~1518) as the product of a massive He-WD+He-WD merger in this evolutionary phase. Here we present a detailed analysis of a second luminous helium-rich O-type hot subdwarf (He-sdO), Gaia~DR3~5369742867090186624 (G5369). We derive an effective temperature of $55~000\pm2500$~K, a surface gravity of $\log g = 4.85\pm0.15$, and an upper limit on the hydrogen abundance of \lognHE$>2.5$. The close similarity of its atmospheric parameters and its helium and carbon abundances to those of HE~1518 indicates that G5369 is also likely the product of a massive He-WD+He-WD merger. In contrast to HE~1518, G5369 is metal-rich, with an iron abundance of about twice solar. The metallicity-adjusted abundance pattern of G5369 for the elements C, N, O, and Ne is broadly consistent with the nucleosynthesis predictions of a composite He-WD merger model and is similar to the observed abundances of R~CrB stars, extreme helium stars, and the O(He) star KPD~0005+5106. The kinematics of G5369 place it near the boundary between the thick disk and halo. Its orbit, with a pericentre of $\sim$3.5 kpc, together with its high metallicity suggests that G5369 most likely originated in the inner Galaxy. 
   }


   \maketitle
   \nolinenumbers
%
%________________________________________________________________
\section{Introduction}
Hot subdwarf stars of spectral type B (sdB) and O (sdO) populate the region below the main sequence (MS) and above the white dwarf (WD) cooling branch in the Hertzsprung-Russell diagram (HRD; for reviews, see \citet{Heber2016,Heber2024}).  Most of them are core or helium shell burning stars with no or very thin hydrogen envelopes. Most hot subdwarfs are proposed to form either through the stripping of a low-mass star ($\lesssim1.8~M_\odot$) at the tip of the red giant branch through interaction with a companion \citep{Han2002, Han2003} or through the merger of two helium WDs \citep{Webbink1984, Zhang2012}. The stripping channel produces an sdB with a canonical mass of about $0.47~M_\odot$, determined by the core mass required to initiate the helium flash. In contrast, He-WD+He-WD mergers produce a broader mass distribution \citep{Dawson2026, Latour2026} and can reach $\sim0.9~M_\odot$, the maximum mass expected from the merger of two helium WDs. 

The stripping of intermediate-mass stars at the end of core hydrogen burning also contributes to the hot subdwarf population. Because helium ignites non-degenerately, the resulting hot subdwarf mass increases with the initial mass of the progenitor \citep{Han2002, ArancibiaRojas2024}. Such systems can therefore produce objects that are both less and more massive than the canonical sdBs, but only a few dozen of higher-mass objects have been discovered despite being predicted in large numbers \citep{Goetberg2018}. Most of these stripped stars have been identified as companions to Be stars \citep{Gies1998, Chojnowski2018, Wang2021, Klement2022, Ramachandran2023, Ramachandran2024, MHorn2026}, including systems presumably observed shortly after mass transfer that are still contracting towards the helium zero-age main sequence (He-ZAMS) \citep[He-ZAMS;][]{Irrgang2020, Bodensteiner2020, ElBadry2022, MHorn2025}. Furthermore, \citet{Dawson2026} and \citet{Latour2026} identified a population of hot subdwarfs that lies below the canonical ($M=0.47$\,$M_\odot$) extreme horizontal branch in the HRD, representing the low-mass end of the intermediate-mass stripping channel. Recently, \citet{Drout2023} discovered ultraviolet (UV)-bright objects below the MS in the Magellanic Clouds, and the analysis by \citet{Goetberg2023} yielded parameters consistent with evolutionary tracks for masses above $1~M_\odot$ for the stripped-star candidates. Two of these objects have optical spectra that closely resemble the luminous helium-rich sdO (He-sdO) HE~1518-0948 (hereafter HE~1518). This star was analysed in the first paper of this series \citep[][hereafter Paper~\textsc{i}]{Pritzkuleit2026}, where it was shown to be the product of a He-WD+He-WD merger and to have a mass of about $0.9~M_\odot$. It is currently in the helium-shell-burning phase and ascending the helium giant branch. Because of their increased luminosities, such stars can mimic more massive stripped stars in photometric selections.

Most He-sdOs have effective temperatures ({\Teff}) between 40 and 55~kK, surface gravities ({\logg}) between 5.5 and 6.0~dex, and stellar radii between 0.2 and $0.3~R_\odot$ \citep[e.g.][]{Stroeer2007, Nemeth2012, Luo2021, Latour2026}. HE~1518 has a similar {\Teff} of about $52~000~\mathrm{K}$ but a lower {\logg} of only 4.6~dex as a result of its inflated radius of about $0.7~R_\odot$. The star has a carbon mass fraction of about 3\% and was most likely formed through a fast merger, during which temperatures became sufficiently high to burn part of the helium into carbon \citep{Zhang2012}. Nitrogen is also enhanced relative to iron due to CNO processing in the progenitors of the helium WDs. However, the optical nitrogen lines are very weak because HE~1518 is ultra metal-poor ($n_\mathrm{Fe}/n_{\mathrm{Fe},\odot}\approx0.025$) and therefore started with very low initial abundances of C, N, and O.

In this paper, we present a detailed analysis of the luminous He-sdO Gaia~DR3~5369742867090186624 (hereafter G5369), which we identified in a survey based on the catalogue of hot subluminous stars compiled by \citet{Culpan2022}. The survey targeted blue ($\mathrm{BP}-\mathrm{RP}<-0.30~\mathrm{mag}$) and bright ($M_G<3.5~\mathrm{mag}$) hot subdwarf candidates, located above the hot subdwarf cloud centred at $M_G\sim4.5~\mathrm{mag}$. G5369 has an apparent $G$ magnitude of $12.75~\mathrm{mag}$ and is located at right ascension $\alpha=174.75522^\circ$ and declination $\delta=-51.34853^\circ$ (J2000). By modelling high-resolution UV and medium-resolution optical spectra, we derived atmospheric parameters and elemental abundances (Sects.~\ref{specan} and \ref{Sect_abu}). These were combined with the available photometric, light curve, and astrometric data to determine the stellar parameters and Galactic orbit of the star (Sects.~\ref{sect:sed}-\ref{kinematics}). Based on these results, we discuss its stellar evolution, Galactic origin, nucleosynthesis, and wind features in comparison to HE~1518 (Sect.~\ref{discussion}).

\section{Observations}
G5369 was observed with X-shooter mounted on the ESO Very Large Telescope (VLT) on 5 January 2023 and 8 July 2023 as part of programmes 110.23TL.001 and 111.24MQ.001, respectively. A third spectrum was obtained on 9 December 2024 with the Space Telescope Imaging Spectrograph (STIS) on board the \textit{Hubble} Space Telescope (HST) as part of programme GO-17697 using the E140M grating. 

X-shooter provides simultaneous spectroscopy in its UVB, VIS, and NIR arms. For the analysis presented in this work, only the UVB and VIS spectra were used, covering the wavelength range 3000-10~000~\AA. The UVB spectra have a resolving power of $R~=~9861$, while the VIS spectra have $R~=~11~333$. The average signal-to-noise ratio is approximately 100 in the first epoch and 175 in the second.

The HST/STIS spectrum covers the UV wavelength range $1144$-$1729~\AA$ at a resolving power of $R~=~45~800$. The target was observed during two consecutive HST orbits, and the individual exposures were co-added, resulting in a signal-to-noise ratio of approximately 35.

\section{Spectral analysis}
\label{specan}

We determined the atmospheric parameters \Teff, \logg, \lognHE, radial velocity (\vrad), and projected rotational velocity (\vsini) from the optical X-SHOOTER spectra by fitting a grid of non-local thermodynamic equilibrium (NLTE) model atmospheres with the Interactive Spectral Interpretation System (ISIS; \citealt{Houck2000}). The model atmospheres were computed with TLUSTY/SYNSPEC \citep{Hubeny2017a} and the atmospheric parameters were determined via $\chi^2$ minimisation. The grid spans effective temperatures between 26~250~K and 65~000~K, surface gravities of $4.5 \le \log g \le 6.75$, and helium abundances of $-1.25~\le$~\lognHE~$\le~4.00$\footnote{Metal line blanketing due to C, N, O, Ne, Si, P, S, and Fe is included at fixed abundances. }. Further details on the modelling and the model grids are given in \citet{Dorsch2024} and Paper~\textsc{i}.

\begin{figure*}
\centering
\includegraphics[width=0.985\textwidth]{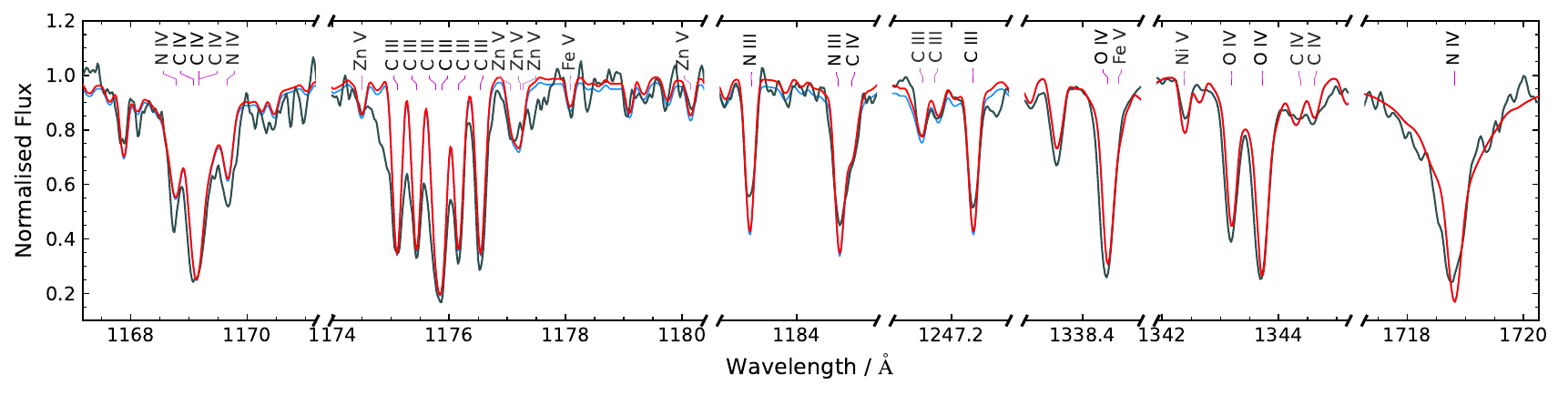}\\[-18pt]
\includegraphics[width=0.985\textwidth]{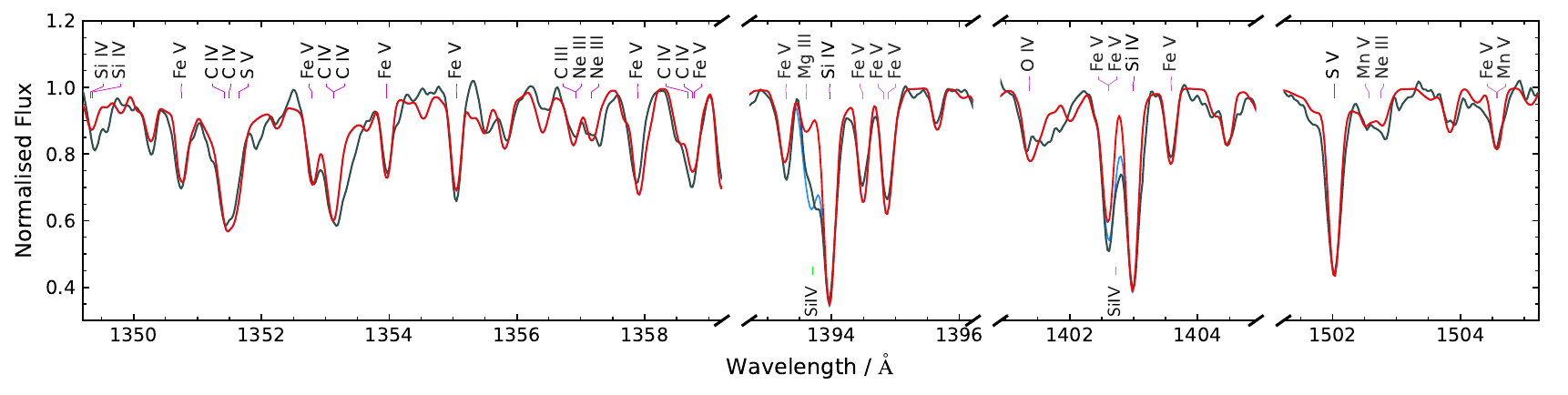}\\[-18pt]
\includegraphics[width=0.985\textwidth]{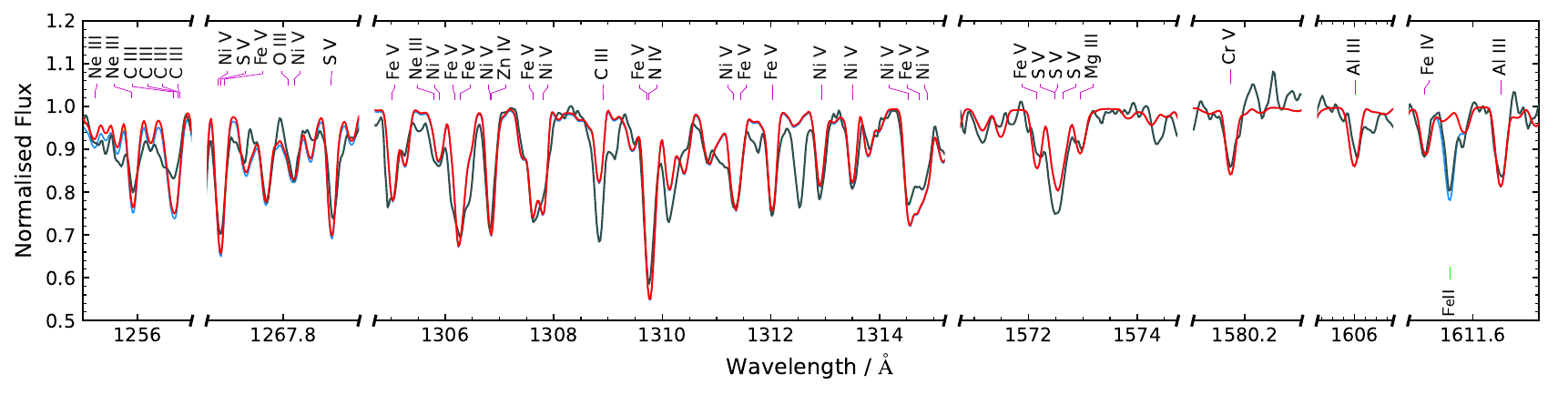}\\[-18pt]
\includegraphics[width=0.985\textwidth]{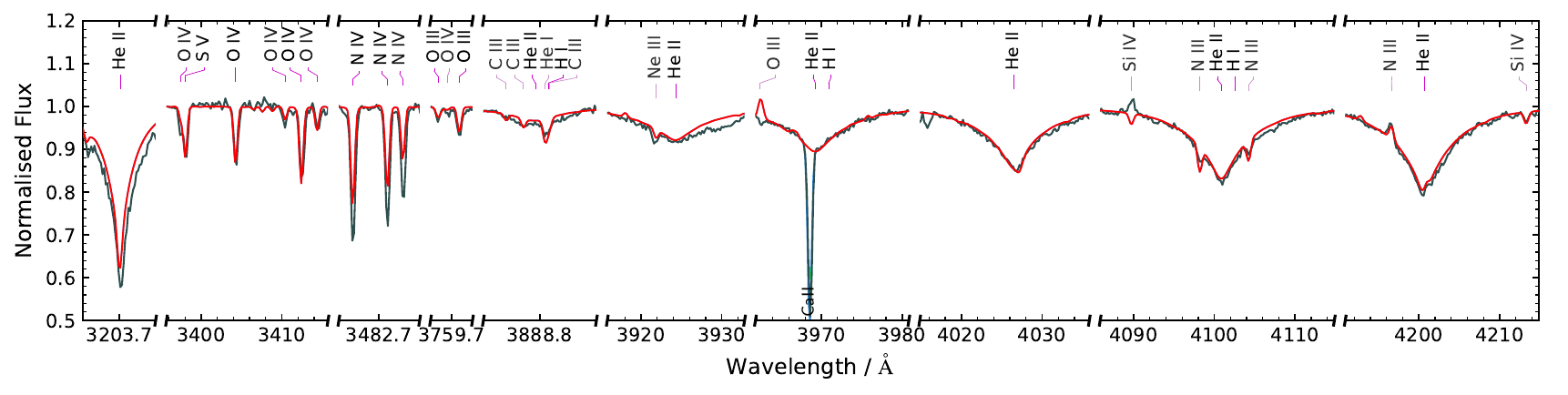}\\[-18pt]
\includegraphics[width=0.985\textwidth]{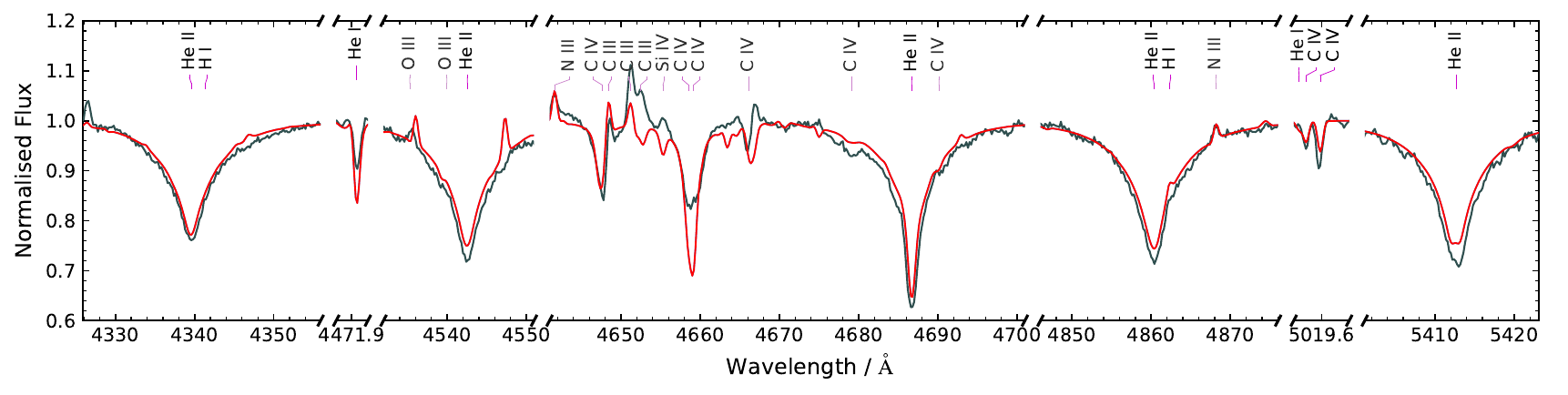}\\[-18pt]
\includegraphics[width=0.985\textwidth]{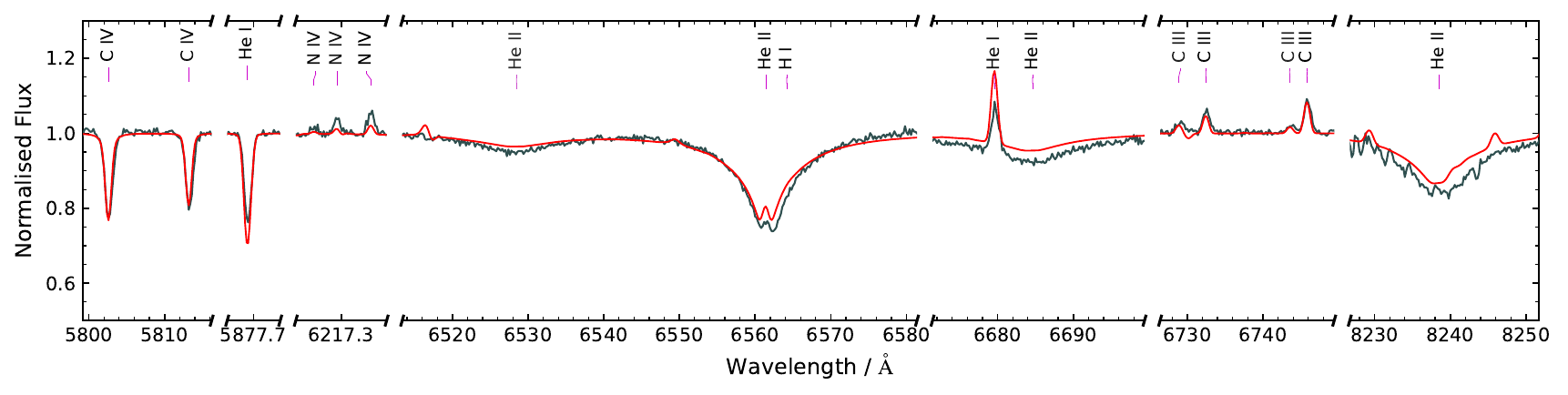}
%\vspace{-5pt}
\caption{Selected UV STIS/E140M (top three rows) and optical X-shooter (bottom three rows) spectral regions (black). The best-fit model is shown in red, with interstellar components in blue. For this figure, the UV spectrum was re-binned to a resolving power of $R=20~000$.}
\label{Specfit}
\end{figure*}

The initial fit resulted in an effective temperature of $57~000~\pm~2500~\mathrm{K}$, a surface gravity of $\log g = 4.85\pm0.15~\mathrm{dex}$, and a projected rotational velocity of $24\pm5~\mathrm{km~s}^{-1}$. As in the case of HE~1518, G5369 is affected by the Balmer/He~\textsc{ii} line problem \citep{Lanz1997, Latour2015} because the model fails to reproduce the observed depths of the He~\textsc{ii} lines, which can be seen in Fig.~\ref{Specfit}. Hydrogen lines are undetected in G5369; we therefore fixed the helium abundance at \lognHE = 3.5, since the $\chi^2$ minimisation otherwise compensates for the mismatch by increasing the hydrogen abundance. Following the procedure described in Paper~\textsc{i}, we derived an upper limit of \lognHE $> 2.5$ for the hydrogen abundance.

\subsection{Abundances}
\label{Sect_abu}

We employed a slightly modified version of the method described in Paper~\textsc{i}. As a first step, we calculated a TLUSTY/SYNSPEC model using the atmospheric parameters {\teff}, {\logg}, and {\lognHE} derived in Sect.~\ref{specan}. The elemental abundances were then adjusted iteratively by eye to obtain an initial estimate until the synthetic spectrum provided a satisfactory match to the observations. Based on this solution, we subsequently computed a SYNSPEC grid in which the abundances were varied by $\pm0.9~\mathrm{dex}$ in steps of $0.3~\mathrm{dex}$. In addition, {\teff} was varied between 54~000 and 60~000~K in steps of 3000~K to account for the influence of the high metal content on the opacities. The elements C, N, O, Ne, Mg, Si, S, Fe, and Ni were treated in NLTE, while all remaining elements were treated in LTE. Finally, a TLUSTY model was recalculated using the updated abundances. The best-fitting model has a temperature of $55~000\pm2500$~K, which is 2000~K lower than the value derived in Sect.~\ref{specan}, but is consistent within the uncertainties. The resulting abundances and their uncertainties are listed in Table~\ref{abundances}, while the comparison of the best-fitting model and the observed spectrum is shown in Fig.~\ref{Specfit}. We adopted the solar abundances from \citet{Asplund2009} as reference values for the fractions listed in Table~\ref{abundances}.

Abundances were determined for C, N, O, Ne, Mg, Al, Si, P, S, Cr, Mn, Fe, Ni, and Zn. Each of these elements has at least one isolated spectral line in the UV range. Although C, N, O, Ne, and Si also show lines in the optical spectrum, many of these transitions originate from highly excited energy levels and are difficult to model accurately. As a result, they are often poorly reproduced by the model spectra. We therefore relied exclusively on the UV spectrum for the abundance determination.

For the intermediate He-sdO $\mathrm{BD}+75^\circ325$, \citet{Lanz1997} reported an effective temperature of $52~000~\mathrm{K}$, similar to that of G5369, and a relatively high microturbulent velocity of $10~\mathrm{km~s^{-1}}$. We found no evidence for such a high microturbulent velocity in G5369 and constrained it to $<5~\mathrm{km~s^{-1}}$.

The elements Mg, Al, P, Cr, Mn, and Zn show only weak lines in the UV spectrum, many of which are blended with Fe and Ni lines. We therefore searched for isolated, uncontaminated lines  to determine their abundances. The selected lines are listed in Table~\ref{elements}. Mg, Al, P, and Zn have fewer than five usable lines. The phosphorus abundance is constrained by a single line and therefore has the largest uncertainty.

In contrast to HE~1518, the UV spectrum of G5369 is rich in Fe and Ni lines. The abundances of C, N, Ne, and Si are enhanced relative to Fe and Ni. While enrichment in C, N, and Ne is predicted by merger models \citep{Zhang2012, Zhang2012b}, the Si enrichment is not (see Sect.~\ref{neon}). The strongest Si feature is the Si~\textsc{iv} 1393.76, 1402.77~{\AA} resonance doublet. The high effective temperature of the star ionises the surrounding interstellar medium (ISM), producing an additional ISM absorption component in these transitions. This does not affect our abundance determination, as the stellar and ISM components are shifted in wavelength due to their different radial velocities, as shown in Fig.~\ref{Specfit}. In addition, the weaker Si~\textsc{iv} lines at 1349.15, 1533.21, 1722.53, and 1727.38~{\AA} are also well reproduced by our spectral fit. The complete UV spectrum is shown in Fig.~\ref{fullspec}. The remaining elements are broadly consistent with the Fe and Ni abundances, which indicate a metallicity of approximately twice solar. 

As in HE~1518, the optical spectrum is dominated by strong He~\textsc{ii} lines and displays the C~\textsc{iii} emission feature at 4650.25~{\AA}, which is bracketed by two C~\textsc{iv} absorption lines. This feature appears to be characteristic of this group of stars. The excellent signal-to-noise ratio of the optical spectrum reveals several weaker features. Among them are the N~\textsc{iii} lines at 4097.32~{\AA} and 4103.39~{\AA} in the wings of the He~\textsc{ii} 4100~{\AA} line and the He~\textsc{i} line at 3888.65~{\AA, which is} visible as a narrow absorption feature within the broad He~\textsc{ii} profile. This feature overlaps with a C~\textsc{iii} triplet. One of its lines at 3889.14~{\AA} is blended with the He~\textsc{i} feature. To determine whether the observed feature is caused by He~\textsc{i} or C~\textsc{iii}, we calculated a test model without the C~\textsc{iii} triplet. The feature remained, confirming that the absorption is primarily caused by He~\textsc{i}.

\begin{table}
\caption{Abundances of G5369 expressed in logarithmic units. }           
\label{abundances}      
\centering                                      
\begin{tabular}{l r r r}          
\toprule\toprule   
Element & number  & mass  & solar fraction\\ 
 & $\log (n_\mathrm{X}/\Sigma_i n_i)$  & $\log (m_\mathrm{X}/\Sigma_i m_i)$  & $\log (n_\mathrm{X}/n_{\mathrm{X},\odot})$\\ \midrule
H &  $<$ -2.41 & $<$ -3.02& $<$ -2.37\\
He &  $-0.005$ & -0.02& 1.10\\
C &  $-2.16\pm0.10$ & $-1.69\pm0.10$& 1.45\\
N &  $-2.63\pm0.20$ & $-2.10\pm0.20$ & 1.57\\
O &  $-3.15\pm0.20$ & $-2.56\pm0.20$& 0.20\\
Ne & $-2.86\pm0.20$ & $-2.17\pm0.20$& 1.25\\
Mg & $-3.80\pm0.30$ &$-3.03\pm0.30$& 0.63\\
Al & $-5.13\pm0.20$ &$-4.31\pm0.20$& 0.46\\
Si & $-3.62\pm0.20$ & $-2.79\pm0.20$ & 0.91\\
P  & $-5.88\pm0.40$ & $-5.01\pm0.40$ & 0.74 \\ 
S &  $-4.40\pm0.20$ & $-3.51\pm0.20$& 0.52\\
Cr & $-5.86\pm0.30$ &$-4.76\pm0.30$& 0.54\\
Mn & $-6.16\pm0.30$ &$-5.04\pm0.30$& 0.45\\
Fe &  $-4.26\pm0.20$ & $-3.13\pm0.20$& 0.27\\  
Ni &$-5.53\pm0.20$ & $-4.38\pm0.20$& 0.29\\
Zn & $-7.10\pm0.20$ &$-5.90\pm0.20$ & 0.38\\
\bottomrule                                           
\end{tabular}
\end{table}

\subsection{Wind features}

\begin{figure}
\centering
\includegraphics[width=1.0\linewidth]{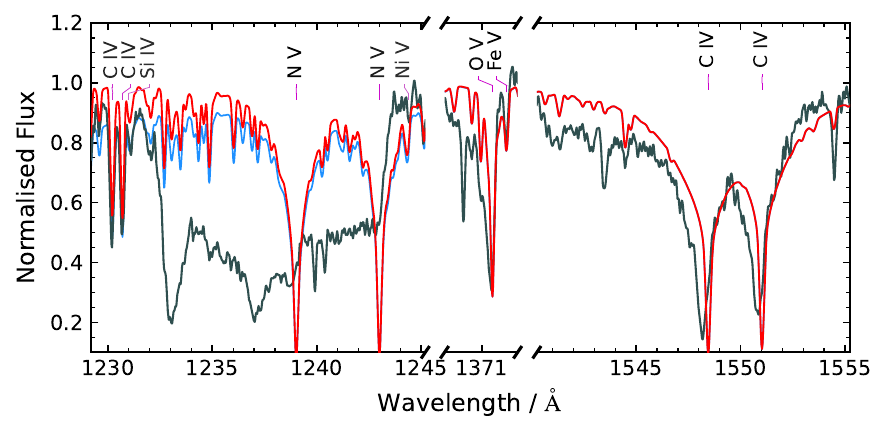}
\caption{UV lines of C, N, and O affected by the stellar wind of G5369. The meaning of the colours is the same as in Fig.~\ref{Specfit}.}
\label{SpecfitWind}
\end{figure}

Some spectral lines are affected by the stellar wind and were excluded from the abundance determination. The most prominent examples are the N~\textsc{v} doublet, the O~\textsc{v} line, the C~\textsc{iv} doublet (Fig.~\ref{SpecfitWind}), and the C~\textsc{iii} sextuplet at 1176~{\AA} (first panel of Fig.~\ref{Specfit}). The C~\textsc{iv} lines appear to be blueshifted, while the N~\textsc{v} resonance doublet shows an unusual profile in which the original doublet is no longer resolved. Instead, it appears as a broad, almost box-shaped absorption feature. Within this structure, two additional absorption components are present whose wavelength separation is identical to that of the N~\textsc{v} doublet. These components likely form in the stellar wind and are shifted by the local wind velocity at the region where they originate. From the N~\textsc{v} doublet we measured a terminal wind velocity of about $1500~\mathrm{km~s}^{-1}$.

Such a line profile has not previously been reported in either O-type stars or hot subdwarfs, where this line typically appears as a P~Cygni profile when the mass-loss rate is sufficiently high. The only comparable case is the He-sdO star CD$-$46~8926 presented by \citet{Husfeld1989} and \citet{Krticka2024}. However, that star has an effective temperature of about 70~000~K, significantly hotter than G5369. Such stars  provide interesting cases to study unusual stellar wind structures, which would contribute to improve our general understanding of mass-loss in hot stars. 

\section{Spectral energy distribution}
\label{sect:sed}
To construct the spectral energy distribution (SED) shown in Fig.~\ref{SED}, we combined the atmospheric parameters {\Teff}, {\logg}, and {\lognHE} derived from the spectroscopic analysis with the \textit{Gaia} DR3 parallax measurement \citep{GaiaDR3}. The parallax was corrected for the zero-point offset following \citet{Lindegren2021}, and its uncertainty was inflated as described by \citet{ElBadry2021}. The free parameters of the fit were the angular diameter, $\theta$, and the colour excess, $E(44-55)$. Interstellar reddening was modelled using the extinction law of \citet{Fitzpatrick2019}. The fit was performed via $\chi^2$ minimisation. We used the same model atmosphere grids as in the spectroscopic analysis. Further details of the fitting procedure can be found in \citet{Heber2018}.

From the SED fit, we derived the stellar radius, luminosity, and mass. The resulting parameters and their $1\sigma$ uncertainties are listed in Table~\ref{params}. Despite the precise \textit{Gaia} parallax measurement, the uncertainty in the mass remains large because it is dominated by the uncertainty in {\logg} (see Sect.~\ref{specan}).

No infrared excess is detected in the SED, ruling out a red-giant or subgiant companion. A low-mass MS companion cannot be excluded, as its flux contribution would be undetectable. However, the absence of significant radial-velocity variations between the three observing epochs makes a close MS companion unlikely. The light curve shown in Fig.~\ref{lightcurve} has a variation of unclear origin, which is discussed in Appendix~\ref{sect_lcurve}.

\begin{table}
\caption{Atmospheric and stellar parameters of G5369. }       
\label{params}      
\centering     
\setstretch{1.1}
\begin{tabular}{l r}
\toprule
\toprule
%Parameter & Value\\
%\midrule
        \multicolumn{2}{c}{Spectral analysis} \\
\midrule
    \Teff~(K) & $55~000\pm~2500$ \\      
    \logg~(cgs) & $4.85~\pm~0.15$ \\
    \lognHE & $>2.5$  \\
    \vsini~(\ensuremath{\mathrm{km~s}^{-1}}) & $24~\pm 5$  \\
    \vrad~(\ensuremath{\mathrm{km~s}^{-1}})& $47~\pm 5$  \\
\midrule
        \multicolumn{2}{c}{SED analysis} \\
\midrule
    \ensuremath{R(55)} & $2.59~\pm~0.08$ \\
    \ensuremath{E(44-55)\,(\mathrm{mag})} & $0.11~\pm~0.02$ \\
    \ensuremath{\log(\theta(\mathrm{rad}})) & $-10.912~\pm~0.009$\\
    \ensuremath{\mathrm{Parallax}\,\varpi} & $0.38~\pm~0.04$\\
    \ensuremath{d_\mathrm{spec}} (kpc) & $2.63^{+0.27}_{-0.22}$ \\[3pt]
    \ensuremath{R/R_\odot} & $0.72^{+0.08}_{-0.07}$ \\[3pt]
    \ensuremath{\log L/L_\odot} & $3.67^{+0.13}_{-0.10}$ \\
    \ensuremath{M/M_\odot} & $1.3^{+0.7}_{-0.5}$ \\
\midrule
        \multicolumn{2}{c}{Evolutionary mass} \\
\midrule
    \ensuremath{M_\mathrm{evol}/M_\odot} & $0.7<M\lesssim0.9$ \\
\midrule
        \multicolumn{2}{c}{Kinematic analysis} \\
\midrule
    \ensuremath{U}~(km~s\ensuremath{^{-1}}) & $180^{+23}_{-18}$\\[3pt]
    \ensuremath{V}~(km~s\ensuremath{^{-1}}) & $169^{+7}_{-7}$\\[3pt]
    \ensuremath{W}~(km~s\ensuremath{^{-1}}) & $-21^{+4}_{-4}$\\ [3pt]
    \ensuremath{e} & $0.58 \pm0.04$ \\ [3pt]
    \ensuremath{L_\mathrm{z}}~(Mpc~km~s$^{-1}$) & $-1.26 \pm 0.09$\\
\bottomrule                                            
\end{tabular}
\end{table}

\section{Galactic kinematics}\label{kinematics}

We computed the kinematic and orbital properties of G5369, using right ascension, declination, proper motions, and parallax from \textit{Gaia} DR3 \citep{GaiaDR3}, with a correction to the parallax as described in Sect.~\ref{sect:sed}, and the radial velocity from our spectroscopic analysis listed in Table~\ref{params}. For the estimation of the Galactic orbital parameters we adopted the barred Milky Way model by \citet{Portail2017} with the analytical approximation given by \citet{Sormani2022} and performed orbital integrations over 5 Gyr saving the orbit's trajectory every 2 Myr.

\begin{figure*}
\centering
\includegraphics[width=0.49\linewidth]{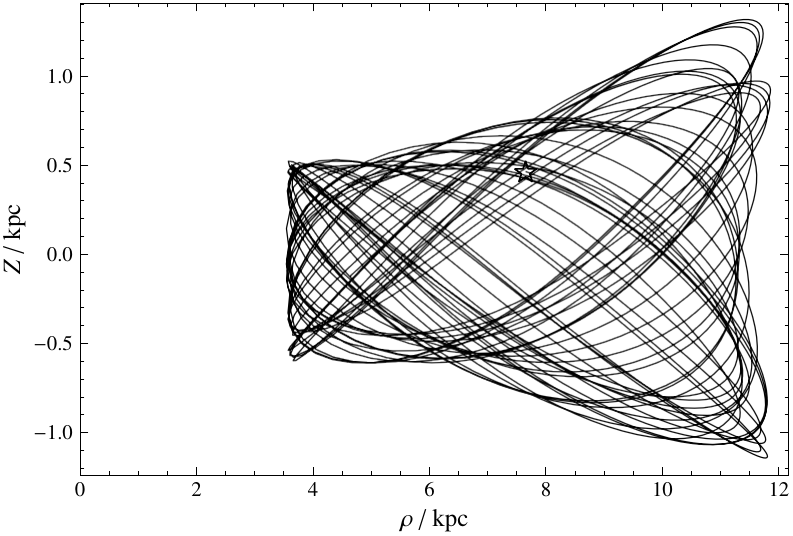}
\includegraphics[width=0.49\linewidth]{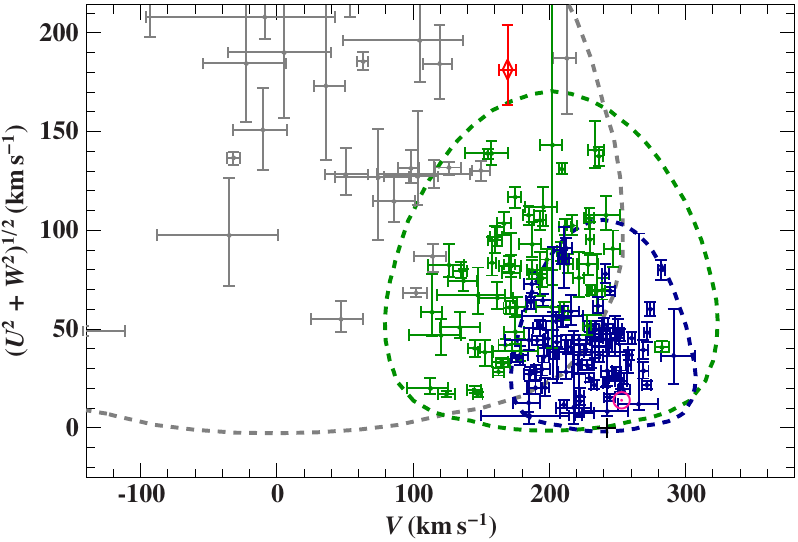}
\caption{\textit{Left}: Galactic radius ($\rho$) vs height ($z$), showing the Galactic orbit and current position of G5369, marked by a star symbol. \textit{Right}: Toomre diagram showing the position of G5369 in red. The dotted blue, green, and grey areas represent the thin disk, thick disk, and halo contours from \citet{Robin2003}. As a comparison, we show the positions of the known He-sdOs from \cite{Dorsch2024}, based on the \citet{Culpan2022} sample. The colour-coding is the same as for the contours and represents their Galactic population membership. The position of the Sun is shown as a pink circle.}
\label{rhoz}
\end{figure*}

Figure~\ref{rhoz} (left) shows the resulting $\rho$--$z$ diagram, where $\rho$ is the Galactocentric distance and $z$ the distance from the Galactic plane. The orbit oscillates between $\rho \simeq 3.5$ and $12~\mathrm{kpc}$ and reaches heights of approximately $\pm1.5~\mathrm{kpc}$. It remains largely confined to the Galactic plane and resembles a thick-disk orbit. The Toomre diagram is shown in Fig.~\ref{rhoz} (right), where the blue, green, and grey contours mark the thin-disk, thick-disk, and halo regions \citep{Robin2003}, respectively. G5369 (red diamond) is located within the halo contour, although its uncertainty overlaps with the thick-disk region. We therefore conclude that G5369 lies near the boundary between the thick disk and halo, an unusual location for a star with such a high metallicity.

\section{Discussion}
\label{discussion}

\subsection{Stellar evolution} \label{HRDdiscussion}

\begin{figure*}
  \includegraphics[width=0.49\linewidth]{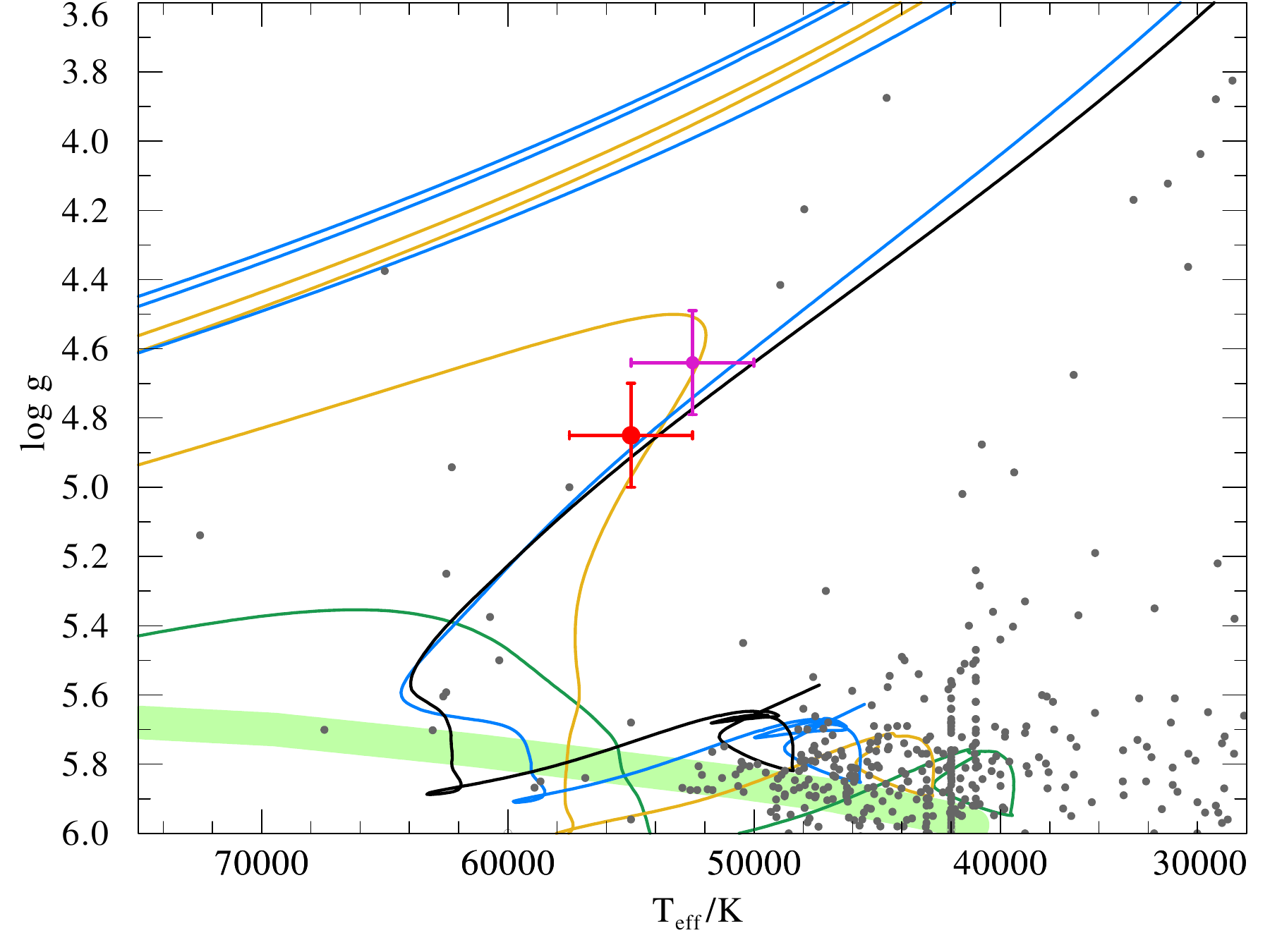}
  \includegraphics[width=0.49\linewidth]{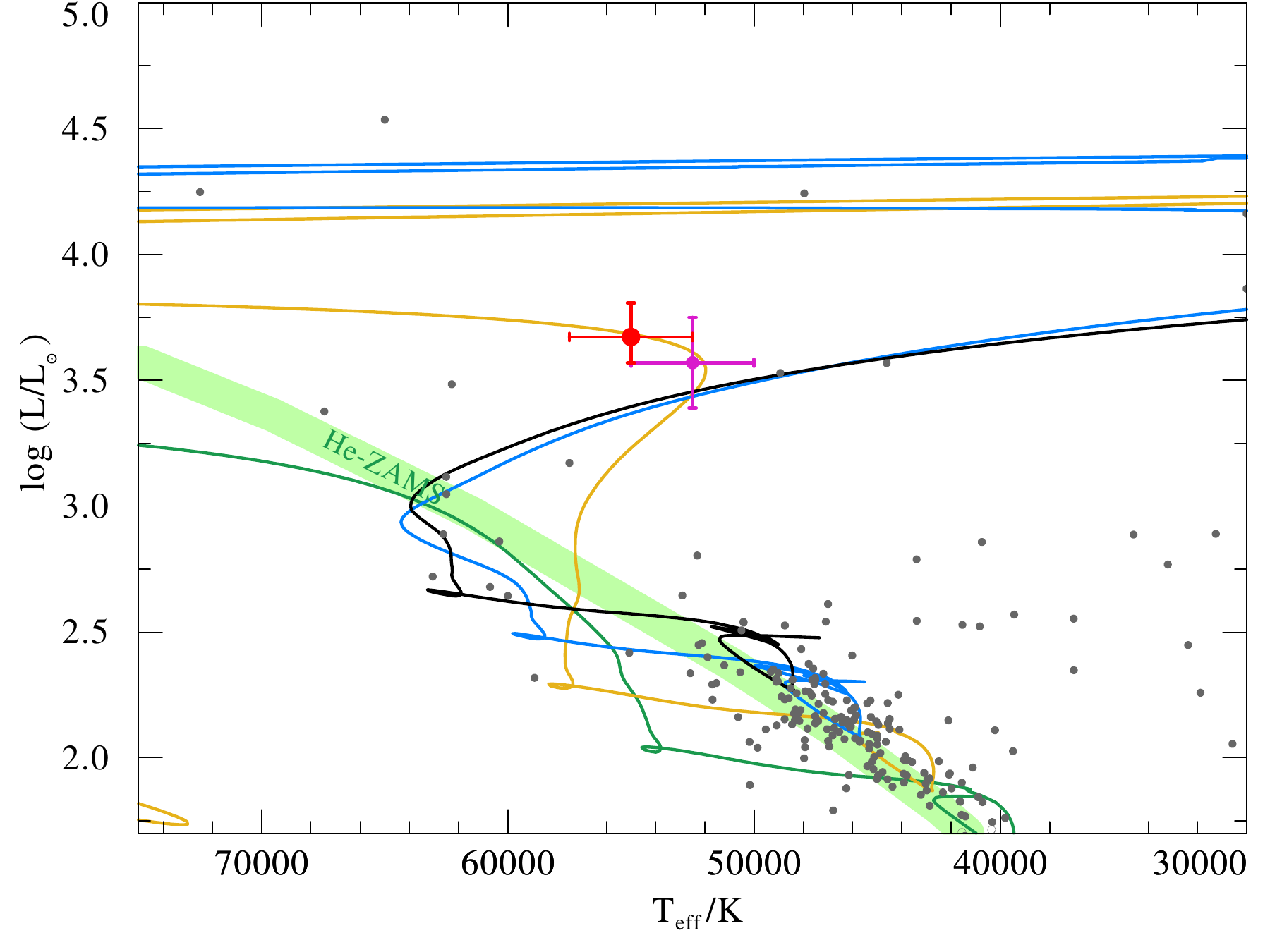}
  \caption{\textit{Left}: Kiel diagram showing the positions of G5369 (red) and HE~1518 (purple). Stellar evolution tracks for helium stars with masses of 0.6 (dark green), 0.7 (gold), 0.8 (blue), and $0.9~M_\odot$ (black) at $Z=0.02$ are shown for comparison. The light green region indicates the location of the He-ZAMS from \citet{Paczynski1971} for helium stars with masses between 0.6 and $\sim$$2.0~M_\odot$. The grey dots are the observed helium-rich hot subdwarfs in the SALT survey of hydrogen-deficient stars \citep[][preliminary results]{Jeffery2026}.
  \textit{Right}: HRD showing the same objects and evolutionary tracks as the Kiel diagram.} 
  \label{Kiel}
\end{figure*}

The majority of helium-rich hot subdwarfs are located on the He-ZAMS, as shown in Fig.~\ref{Kiel}. Most are thought to be the outcome of He-WD+He-WD mergers. In Paper~\textsc{i}, we showed that the luminous He-sdO HE~1518 is also most likely formed through this channel, with the two merging He-WDs having masses close to the maximum possible for He-WDs. Helium stars with masses above $\sim0.7~M_\odot$ do not evolve directly towards the WD cooling sequence but instead expand after core helium depletion in the core and the ignition of helium-shell burning.

The close similarity in atmospheric parameters and carbon abundance between G5369 and HE~1518 suggests that G5369 might also be the product of a He-WD+He-WD merger. In Fig.~\ref{Kiel}, we show the position of G5369 (red) in the {\teff}-{\logg} diagram (left) and in the HRD (right). For comparison, we also include HE~1518 (purple) and the helium-rich hot subdwarfs from the Southern African Large Telescope (SALT) survey of hydrogen-deficient stars (grey) \citep{Jeffery2026}. In both diagrams, G5369 is located close to HE~1518. We also show a set of metal-rich ($Z=0.02$) helium-star evolutionary tracks with masses of $0.6~M_\odot$ (dark green), $0.7~M_\odot$ (gold), $0.8~M_\odot$ (blue), and $0.9~M_\odot$ (black), introduced in Paper~\textsc{i}. The $0.6~M_\odot$ track does not reach the position of G5369, indicating that the star is more massive. In both the {\teff}-{\logg} diagram and the HRD, the parameters of G5369 are broadly consistent with evolutionary tracks for merger products with masses between $0.7$ and $0.9~M_\odot$. However, none of these tracks reproduces the observed position in both diagrams at the same evolutionary stage, preventing a more precise mass estimate. The total mass is expected to remain below $\sim0.9~M_\odot$, the maximum possible mass for a He-WD+He-WD merger. This mass range is also consistent with the mass derived from the SED and parallax. 

Depending on its current evolutionary state, G5369 may evolve towards a cooler helium giant phase, where it would resemble an extreme helium (EHe) star and, at even lower temperatures, as a hydrogen-deficient carbon (HdC) star, possibly even as a dusty  R CrB star \citep{Zhang2012b}. The star would then evolve back towards higher temperatures, passing through a more luminous EHe phase, before reaching the O(He) stage \citep{Reindl2014} and ultimately entering the WD cooling sequence.

\subsection{Potential ejection mechanisms}
\label{Galbareje}

The thick disk and halo are predominantly composed of old, metal-poor stellar populations. Star formation in these components ceased before sufficient chemical enrichment could produce the metallicities observed in G5369. \citet{Crawford2025} identified red giants with masses exceeding $2~M_\odot$ on halo orbits. Such objects cannot be explained by mergers of stars formed in the halo, as the present-day turn-off mass of $\approx0.8~M_\odot$ limits the maximum merger mass to $\approx1.6~M_\odot$. They therefore concluded that these stars must have been ejected from the Galactic disk. Similarly, \citet{Dantas2023} identified a population of old metal-rich stars with metallicities comparable to that of G5369 on thick-disk and halo-like orbits. These objects are thought to originate in the inner Galaxy, where the early higher star-formation rate led to more efficient chemical enrichment than in the solar neighbourhood (see e.g. \citealt{Barbuy2018} for a review).

Several mechanisms for ejecting stars from the Galactic disk have been proposed. One possibility is ejection during the hot-subdwarf phase by a type Ia supernova, in which a massive C/O WD accretes helium-rich material from a Roche-lobe-filling hot subdwarf companion \citep{Rajamuthukumar2025}. Because such binaries are extremely compact, the hot subdwarf is expected to be ejected with velocities between $\sim500~\mathrm{km,s^{-1}}$ and $1200~\mathrm{km,s^{-1}}$ \citep{Neunteufel2022}. These velocities far exceed the inferred ejection velocity of $171^{+8}_{-7}~\mathrm{km,s^{-1}}$ for G5369.  US~708 is the only known hot subdwarf whose properties are consistent with ejection by a Type Ia supernova \citep{Geier2015}. In contrast to G5369, it is enriched in nitrogen but carbon deficient. We therefore reject a Type Ia supernova as the origin of G5369.

Another mechanism was proposed by \citet{Gao2019}, in which the progenitor binary is initially part of a hierarchical triple system. In this scenario, the outer companion is a massive star that eventually explodes as a supernova, ejecting the remaining binary with velocities up to $400~\mathrm{km~s^{-1}}$.

Another possibility in addition to the supernova-driven scenarios discussed above are dynamical interactions in stellar birth clusters, involving encounters either between a single star and a binary or between two binaries. In this scenario, the  progenitor system of G5369 was likely ejected as an intact binary before the He-WD merger that produced the hot subdwarf, given that the birth cluster would have dispersed before the MS progenitors had time to evolve and interact. 
Because binaries become increasingly rare at higher ejection velocities \citep{Kroupa1998, Perets2012}, the inferred velocity of G5369 disfavours this scenario. Moreover, binaries ejected at such high velocities are expected to be very compact. In fact, studies of B-type runaway binaries suggest that close systems are preferentially ejected \citep{Chen2025}. Assuming that the progenitor of G5369 consisted of two solar-type stars of nearly equal mass, the primary must have expanded sufficiently to initiate mass transfer and subsequently form a $0.3$--$0.4~M_\odot$ helium WD. From MIST stellar evolutionary tracks \citep{MIST1, MIST2} we find that the stellar radius is expected to be $\sim20$--$40~R_\odot$ at this stage for progenitor masses between 1.8 and $0.8~M_\odot$, respectively, implying Roche-lobe overflow in binaries with orbital periods of roughly 20--100 days. Such wide systems are unlikely to survive the close dynamical encounter required to reach the inferred ejection velocity. We therefore consider this scenario unlikely.

While the scenarios above focus on the dynamical history of G5369 as an individual object, its metallicity and kinematics closely resemble the old metal-rich population identified by \citet{Dantas2023}. Using a sample of similar stars as tracers, \citet{Nepal2024} showed that they were most likely redistributed from the inner Galaxy by the formation of the Galactic bar approximately 3~Gyr ago. These populations should naturally contain progenitor binaries capable of producing stars like G5369, without requiring a specialised ejection mechanism. In Fig.~\ref{kinematics_corot}, we show a projection of the stellar orbit onto the $X$-$Y$ plane in the co-rotating frame. As can be seen, the orbit of G5369 forms loops in the inner regions close to the Galactic bar, which may indicate that the star is located near a Lagrange point of the bar. Stars with similar orbital behaviour were identified by \citet{PerezVillegas2017}. We therefore argue that G5369 likely belongs to this redistributed inner-Galaxy population. However, stars like G5369 cannot be utilised to determine the age of the Galactic bar due to their uncertain merger time and masses of the progenitor stars.

\subsection{Nucleosynthesis}\label{neon}

\begin{figure*}
  \includegraphics[width=0.49\linewidth]{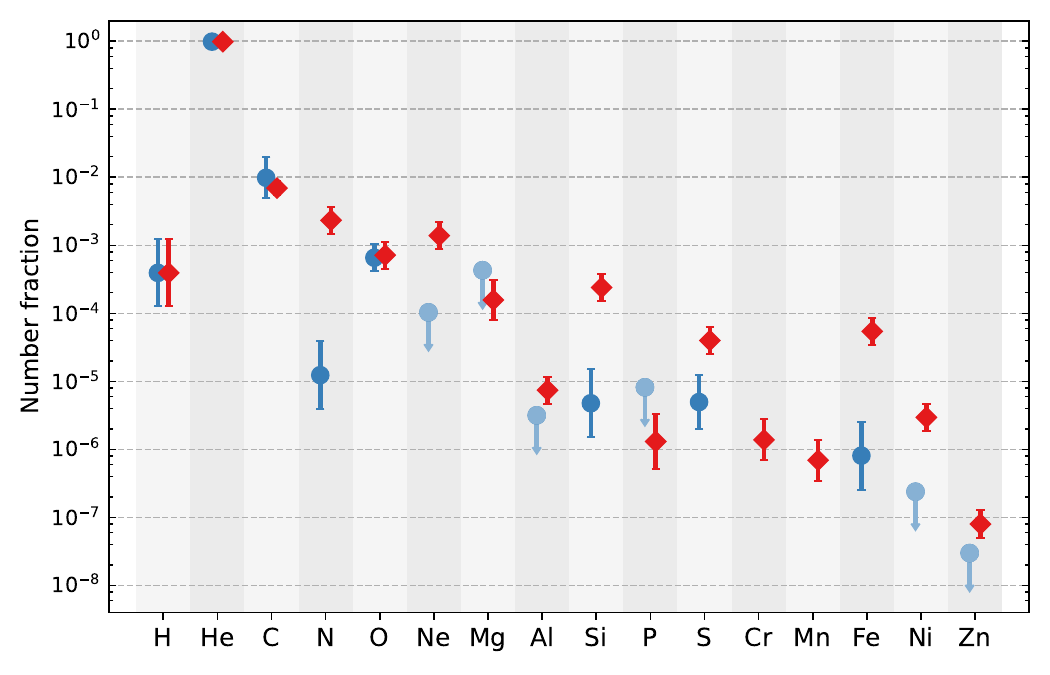}
  \includegraphics[width=0.49\linewidth]{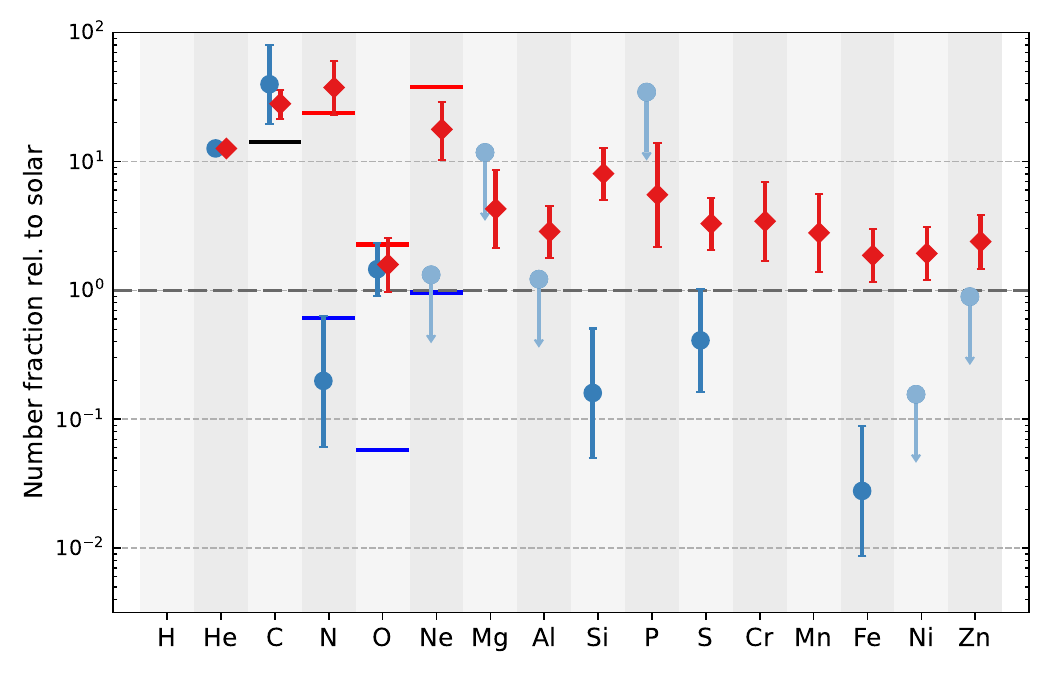}
  \caption{\textit{Left}: Number fraction of selected elements of G5369 (red diamonds) compared to HE~1518 (blue points). The downward pointing arrows indicate the upper limits for HE~1518. 
  \textit{Right}: Same as the left plot, but the number fractions are shown relative to solar values. The plot also includes the predicted abundances of C, N, O, and Ne from the composite merger model of \citet{Zhang2012}, with the N, O, and Ne abundances adjusted for the metallicities of G5369 (red bars) and HE~1518 (blue bars). The same predicted C abundance is shown for both stars (black bar).  }
  \label{comp_abu}
\end{figure*}

\begin{figure*}
  \includegraphics[width=0.99\textwidth]{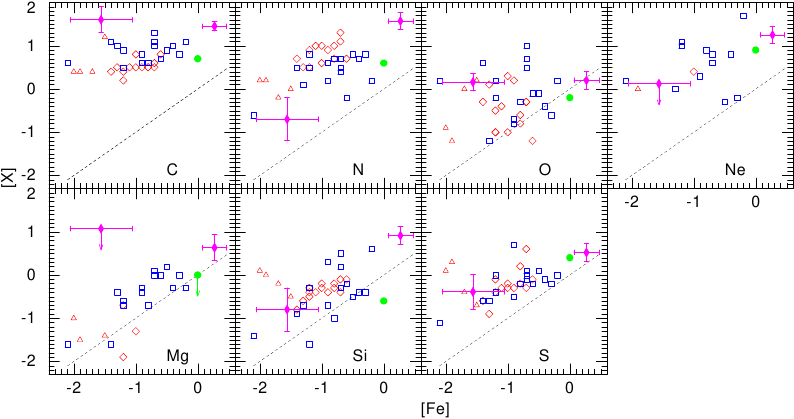}
  \caption{Observed surface abundances as a function of the iron abundance, with all abundances expressed relative to solar values, for majority (red diamonds) and minority (red triangles) R~CrB stars, EHe stars (blue squares) from \citet[and references therein]{Jeffery2011}, the O(He) star KPD~0005+5106 (green circle) \citep{Werner2015}, and the luminous He-sdOs HE~1518 (from Paper~\textsc{i}) and G5369 (pink rhombus with error bars). Upper limits are indicated by downward pointing arrows. The dashed line shows the solar composition scaled to the iron abundance. }
  \label{abu_mergers}
\end{figure*}

Before the merger, the progenitor He-WDs are expected to have undergone CNO processing, during which most of the initial carbon and oxygen was converted into nitrogen. In the composite merger model of \citet{Zhang2012,Zhang2012b}, $^{18}\mathrm{O}$ is produced through $\alpha$-capture on $^{14}\mathrm{N}$, initially forming $^{18}\mathrm{F}$, which subsequently decays to $^{18}\mathrm{O}$. Further $\alpha$-capture on $^{18}\mathrm{O}$ then produces \neon. G5369 is enriched in C, N, and Ne, and its abundance pattern agrees well with the predictions of the composite merger model from \citet{Zhang2012} (red bars) shown in Fig.~\ref{comp_abu} (right). In this figure, we also compare the predicted abundances with those of HE~1518. When the abundance predictions of \citet{Zhang2012} are scaled to the metallicity of HE~1518 (blue bars), they remain in good agreement with its C and N abundances. Its oxygen abundance, however, is substantially higher than predicted and is strikingly similar to that of G5369. Their model predicts that the oxygen produced during the merger originates exclusively from the $\alpha$-capture sequence starting from $^{14}\mathrm{N}$ and not from $\alpha$-capture on carbon. The similar oxygen abundances of G5369 and HE~1518 therefore suggest that an additional source of oxygen might be required at low metallicities.

In Fig.~\ref{abu_mergers}, we compare the abundances of C, N, O, Ne, Mg, Si, and S with measurements of R~CrB and EHe stars \citep{Jeffery2011}, as well as the O(He) star KPD~0005+5106 \citep{Werner2015}. These stars are also considered possible products of WD mergers, although some of them can also result from mergers between a C/O-WD and a He-WD. The dotted line in each panel of Fig.~\ref{abu_mergers} indicates the solar abundance scaled with metallicity. The first row shows elements expected to be affected by the merger, and the second row shows elements that are not expected to be affected. Carbon shows no clear trend with metallicity. This is consistent with most of the carbon being produced by the triple-$\alpha$ process during the merger. Nitrogen is enriched and appears to scale with metallicity, suggesting that most of it originates from CNO-processed material. For HdC stars, \citet{Mehla2025} found that approximately 8\% of the nitrogen is converted into $^{18}\mathrm{O}$ during their formation. This fraction may not apply to He+He WD mergers, as the composite merger models of \citet{Zhang2012} show that the conversion depends on the final mass of the merger product. The oxygen abundances show a different behaviour. While some objects seem to follow the metallicity-scaled solar abundance, others are scattered above the solar-scaled line and have oxygen abundances similar to those of G5369 and HE~1518. In R~CrB and EHe stars, such excesses may be explained by C/O-rich material mixed up from the C/O core of the accretor in addition to $^{18}$O from $\alpha$-captures on $^{14}$N \citep{Jeffery2011}. However, this does not apply to HE~1518. As a double He-WD merger product, it has no C/O core from which oxygen-rich material could have been dredged up. Moreover, even converting all of its nitrogen into $^{18}$O would not be enough to explain the observed oxygen abundance. This suggests that the observed oxygen excess could be due to $^{16}$O produced by $\alpha$-captures on $^{12}$C during the merger. However, the composite merger model of \citet{Zhang2012} does not predict the production of additional $^{16}$O, and further modelling is needed to investigate whether this process could account for the higher oxygen abundance. 

{\neon} plays a crucial role in delaying WD cooling. Evidence for this effect was first found in the metal-rich open cluster NGC~6791 \citep{Althaus2010} and later for field WDs in the 100~pc colour-magnitude diagram from \textit{Gaia} Data Release~2 \citep{GaiaDR2HRD2018}, where a population of ultra-massive ($M>1.05~M_\odot$) WDs create an overdensity known as the Q branch. These are thought to be predominantly C/O WDs \citep{Camisassa2019, Bauer2020, Camisassa2022}, where sedimentation of {\neon} and other neutron-rich isotopes releases gravitational energy that significantly delays cooling \citep{Blouin2021, Bedard2024}.

The cooling delay is not restricted to ultra-massive WDs. In their study of the metal-rich open cluster NGC~6791, \citet{Salaris2024} found that WDs of lower mass can also experience cooling delays lasting several Gyr. NGC~6791 has a reported metallicity of $\left[\mathrm{Fe}/\mathrm{H}\right]=0.29\pm0.10$~dex (where $Z_\odot=0.012$; \citealt{Brogaard2011, Brogaard2021}), which is comparable to what we find for G5369. The currently observed {\neon} abundance in G5369 ($0.67^{+0.20}_{-0.12}$~\% by mass) is too low to produce a significant cooling delay. However, if the current photospheric abundance is representative of the bulk composition of the envelope above the helium-burning shell, then the nitrogen produced by CNO processing in the progenitor stars is expected to be converted into {\neon} through $\alpha$-capture during both core and shell helium burning. If all of this nitrogen were converted, the {\neon} mass fraction would increase to $2.00^{+0.83}_{-0.58}~\%$. This is only slightly below the $\sim$2.5--3\% {\neon} mass fraction found by \citet{Bedard2024} to reproduce the cooling delay of Q-branch WDs. Whether such an abundance is sufficient to produce a significant cooling delay is currently uncertain. The cooling models of \citet{Bedard2024} show that the predicted delay depends on how {\neon} distillation is treated when its abundance approaches the threshold at which the solid phase becomes no longer buoyant. Thus, a final {\neon} abundance of $2.00^{+0.83}_{-0.58}~\%$ does not allow us to determine whether G5369 will experience a substantial cooling delay, but places it within the abundance regime where such a delay may be possible.

\citet{Cheng2019} used the age-velocity-dispersion relation to argue that WDs on the Q branch experience long cooling delays, based on their large transverse velocities. G5369 has a transverse velocity of $276^{+23}_{-28}~\mathrm{km~s}^{-1}$ and would satisfy this criterion. However, as discussed in Sect.~\ref{Galbareje}, its orbit was likely altered by interactions with the Galactic bar, so this relation may not apply. 

The observed overabundance of silicon could be explained by $\alpha$-enhancement typical of old metal-poor stellar populations, but this signature is not observed in old metal-rich stars in the solar neighbourhood \citep{Dantas2023}. In G5369, the magnesium and sulphur abundances are consistent with solar-scaled abundances within the uncertainties and do not show a comparable overabundance. Similar overabundances of silicon of unclear origin, reaching up to one dex above the metallicity-scaled abundance, have also been observed in some R~CrB and EHe stars \citep{Jeffery2011}.
If silicon was not enriched in the primordial material that formed the progenitor stars it is difficult to explain the origin. A possible production channel may be further $\alpha$-captures on {\neon}, forming the neutron-rich isotope $^{30}\mathrm{Si}$. However, the temperatures reached during the merger are not high enough for efficient silicon production, making this scenario unlikely.

Alternatively, the observed silicon abundance might result from accretion of rocky material from a circumstellar debris disk that could have formed during the merger. Studies of polluted WDs show that accreted refractory material is rich in Mg, Si, and Fe, although the relative abundances of these elements vary between systems \citep[and references therein]{Cutolo2026, Williams2024}. Further investigation is needed to assess whether circumstellar disks around hot subdwarfs like G5369 are able to form rocky bodies that can then be re-accreted onto the star.

\section{Summary and conclusions}

In this paper, we presented the second detailed analysis of a hot subdwarf ascending the helium giant branch. We find that G5369 has stellar parameters similar to those of HE~1518 \citep{Pritzkuleit2026} and conclude that it is also the product of a He-WD+He-WD merger. In contrast to the extremely metal-poor halo star HE~1518 analysed in Paper~\textsc{i}, G5369 has a super-solar metallicity despite moving on an orbit at the transition between the thick disk and halo, where such metal-rich stars are not expected to form. 
Our kinematic analysis suggests that G5369 most likely originated in the inner, more metal-rich part of the Galaxy. Its orbit is similar to those of metal-rich MS stars for which \citet{Nepal2024} showed evidence of redistribution from the inner Galaxy by the Galactic bar.

During the merger, temperatures were sufficiently high to convert part of the helium into carbon, while part of the nitrogen produced through CNO processing in the progenitor stars was converted into {\neon}. Our C, N, O, and Ne abundance measurements agree well with the predicted pattern of the composite merger model of \citet{Zhang2012}. During the subsequent core and helium-shell-burning phases, further $\alpha$-capture on the remaining nitrogen may produce additional {\neon}. Assuming complete conversion of the remaining nitrogen, G5369 with an initial metallicity of approximately twice solar (where {$Z_\odot=0.014$}) could reach a final {\neon} mass fractions of $2.00^{+0.83}_{-0.58}~\%$.

This {\neon} mass fraction is close to the $\sim$2.5--3\% {\neon} mass fraction found by \citet{Bedard2024} to reproduce the cooling delay of Q-branch WDs. Whether G5369 will experience a substantial cooling delay remains uncertain because of uncertainties in current models for the threshold for {\neon} distillation.

Luminous He-sdOs also exhibit signatures of stellar winds in their UV spectra. G5369 is particularly interesting because it displays line profiles unlike those previously observed in O-type or hot subdwarf stars, suggesting that its wind structure is different. The luminous He-sdOs may provide an opportunity to investigate the winds of intermediate-mass stripped stars at various metallicities. \citet{Goetberg2017} suggest that intermediate-mass stripped stars could have played an important role in the re-ionisation of helium in the early Universe, but their He~\textsc{ii}-ionising photon flux is highly sensitive to the adopted mass-loss rates. The two currently known objects, HE~1518 and G5369, probe the extremely metal-poor and super-solar regimes, respectively, suggesting that a larger sample of luminous He-sdOs could be used to investigate the metallicity dependence of their winds. 
Luminous He-sdOs thus offer a significant observational advantage because they are much closer and therefore brighter than potential stripped stars in extremely metal-poor galaxies, making detailed UV spectroscopy achievable with HST. 

HE~1518 and G5369 provide direct observational constraints on the outcome of He-WD+He-WD mergers at opposite metallicity extremes, particularly on the evolution beyond core helium burning, and on nucleosynthesis during the hot merger phase.

\begin{acknowledgements}
MD was supported by the Deutsches Zentrum für Luft- und Raumfahrt (DLR) through grant 50-OR-2510. 
MP received funding by the Deutsche Forschungs-
gemeinschaft (DFG) through grants GE2506/9-1, GE2506/12-1, and GE2506/18-1. HD was supported by the DFG through grant  GE2506/9-2.
N.R. is supported by the Deutsche Forschungsgemeinschaft (DFG) through grant RE3915/2-1.

Part of this work was supported by the
CONICET-DAAD 2022 bilateral cooperation grant number 80726. M3B is partially funded by CONICET and Agencia I+D+i through grants PIP-2971 and
PICT 2020-03316. 

Based on observations made with the \textit{Isaac Newton} and \textit{William Herschel} telescopes operated on the island of La Palma by the Isaac Newton Group of Telescopes in the Spanish Observatorio del Roque de los Muchachos of the Instituto de Astrofísica de Canarias.

Based on observations made with ESO Telescopes at the La Silla Paranal Observatory under programme IDs 167.D-0407(A) and 71.D-0380(A).

Based on observations obtained at the Southern Astrophysical Research (SOAR) telescope, which is a joint project of the Minist\'{e}rio da Ci\^{e}ncia, Tecnologia e Inova\c{c}\~{o}es (MCTI/LNA) do Brasil, the US National Science Foundation’s NOIRLab, the University of North Carolina at Chapel Hill (UNC), and Michigan State University (MSU).

This research is based on observations made with the NASA/ESA Hubble Space Telescope obtained from the Space Telescope Science Institute, which is operated by the Association of Universities for Research in Astronomy, Inc., under NASA contract NAS 5-26555. These observations are associated with program A far-UV treasury survey for hot subdwarf stars (ID 17697; PI: M.Dorsch)
\end{acknowledgements}

%-------------------------------------------------------------------
\bibliographystyle{aa}
\bibliography{sample}

@ARTICLE{Heber2016,
       author = {{Heber}, U.},
        title = "{Hot Subluminous Stars}",
      journal = {\pasp},
         year = 2016,
        month = aug,
       volume = {128},
       number = {966},
        pages = {082001},
          doi = {10.1088/1538-3873/128/966/082001},
archivePrefix = {arXiv},
       eprint = {1604.07749},
 primaryClass = {astro-ph.SR},
       adsurl = {https://ui.adsabs.harvard.edu/abs/2016PASP..128h2001H}
}

@ARTICLE{Goetberg2018,
       author = {{G{\"o}tberg}, Y. and {de Mink}, S.~E. and {Groh}, J.~H. and {Kupfer}, T. and {Crowther}, P.~A. and {Zapartas}, E. and {Renzo}, M.},
        title = "{Spectral models for binary products: Unifying subdwarfs and Wolf-Rayet stars as a sequence of stripped-envelope stars}",
      journal = {\aap},
         year = 2018,
        month = jul,
       volume = {615},
          eid = {A78},
        pages = {A78},
          doi = {10.1051/0004-6361/201732274},
archivePrefix = {arXiv},
       eprint = {1802.03018},
 primaryClass = {astro-ph.SR},
       adsurl = {https://ui.adsabs.harvard.edu/abs/2018A&A...615A..78G}
}

@ARTICLE{Neunteufel2022,
       author = {{Neunteufel}, P. and {Preece}, H. and {Kruckow}, M. and {Geier}, S. and {Hamers}, A.~S. and {Justham}, S. and {Podsiadlowski}, Ph.},
        title = "{Properties and applications of a predicted population of runaway He-sdO/B stars ejected from single degenerate He-donor SNe}",
      journal = {\aap},
         year = 2022,
        month = jul,
       volume = {663},
          eid = {A91},
        pages = {A91},
          doi = {10.1051/0004-6361/202142864},
archivePrefix = {arXiv},
       eprint = {2112.07469},
 primaryClass = {astro-ph.SR},
       adsurl = {https://ui.adsabs.harvard.edu/abs/2022A&A...663A..91N}
}

@ARTICLE{Geier2015,
       author = {{Geier}, S. and {F{\"u}rst}, F. and {Ziegerer}, E. and {Kupfer}, T. and {Heber}, U. and {Irrgang}, A. and {Wang}, B. and {Liu}, Z. and {Han}, Z. and {Sesar}, B. and {Levitan}, D. and {Kotak}, R. and {Magnier}, E. and {Smith}, K. and {Burgett}, W.~S. and {Chambers}, K. and {Flewelling}, H. and {Kaiser}, N. and {Wainscoat}, R. and {Waters}, C.},
        title = "{The fastest unbound star in our Galaxy ejected by a thermonuclear supernova}",
      journal = {Science},
         year = 2015,
        month = mar,
       volume = {347},
       number = {6226},
        pages = {1126-1128},
          doi = {10.1126/science.1259063},
archivePrefix = {arXiv},
       eprint = {1503.01650},
 primaryClass = {astro-ph.SR},
       adsurl = {https://ui.adsabs.harvard.edu/abs/2015Sci...347.1126G}
}

@ARTICLE{Heber2018,
       author = {{Heber}, Ulrich and {Irrgang}, Andreas and {Schaffenroth}, Johannes},
        title = "{Spectral energy distributions and colours of hot subluminous stars}",
      journal = {Open Astronomy},
         year = 2018,
        month = feb,
       volume = {27},
       number = {1},
        pages = {35-43},
          doi = {10.1515/astro-2018-0008},
archivePrefix = {arXiv},
       eprint = {1712.06546},
 primaryClass = {astro-ph.SR},
       adsurl = {https://ui.adsabs.harvard.edu/abs/2018OAst...27...35H}
}

@ARTICLE{Fitzpatrick2019,
       author = {{Fitzpatrick}, E.~L. and {Massa}, Derck and {Gordon}, Karl D. and {Bohlin}, Ralph and {Clayton}, Geoffrey C.},
        title = "{An Analysis of the Shapes of Interstellar Extinction Curves. VII. Milky Way Spectrophotometric Optical-through-ultraviolet Extinction and Its R-dependence}",
      journal = {\apj},
         year = 2019,
        month = dec,
       volume = {886},
       number = {2},
          eid = {108},
        pages = {108},
          doi = {10.3847/1538-4357/ab4c3a},
archivePrefix = {arXiv},
       eprint = {1910.08852},
 primaryClass = {astro-ph.GA},
       adsurl = {https://ui.adsabs.harvard.edu/abs/2019ApJ...886..108F}
}

@ARTICLE{Hubeny2017a,
       author = {{Hubeny}, Ivan and {Lanz}, Thierry},
        title = "{A brief introductory guide to TLUSTY and SYNSPEC}",
      journal = {arXiv e-prints},
         year = 2017,
        month = jun,
          eid = {arXiv:1706.01859},
        pages = {arXiv:1706.01859},
          doi = {10.48550/arXiv.1706.01859},
archivePrefix = {arXiv},
       eprint = {1706.01859},
 primaryClass = {astro-ph.SR},
       adsurl = {https://ui.adsabs.harvard.edu/abs/2017arXiv170601859H}
}

@ARTICLE{Han2002,
       author = {{Han}, Z. and {Podsiadlowski}, Ph. and {Maxted}, P.~F.~L. and {Marsh}, T.~R. and {Ivanova}, N.},
        title = "{The origin of subdwarf B stars - I. The formation channels}",
      journal = {\mnras},
         year = 2002,
        month = oct,
       volume = {336},
       number = {2},
        pages = {449-466},
          doi = {10.1046/j.1365-8711.2002.05752.x},
archivePrefix = {arXiv},
       eprint = {astro-ph/0206130},
 primaryClass = {astro-ph},
       adsurl = {https://ui.adsabs.harvard.edu/abs/2002MNRAS.336..449H}
}

@ARTICLE{Han2003,
       author = {{Han}, Z. and {Podsiadlowski}, Ph. and {Maxted}, P.~F.~L. and {Marsh}, T.~R.},
        title = "{The origin of subdwarf B stars - II}",
      journal = {\mnras},
         year = 2003,
        month = may,
       volume = {341},
       number = {2},
        pages = {669-691},
          doi = {10.1046/j.1365-8711.2003.06451.x},
archivePrefix = {arXiv},
       eprint = {astro-ph/0301380},
 primaryClass = {astro-ph},
       adsurl = {https://ui.adsabs.harvard.edu/abs/2003MNRAS.341..669H}
}

@ARTICLE{Webbink1984,
       author = {{Webbink}, R.~F.},
        title = "{Double white dwarfs as progenitors of R Coronae Borealis stars and type I supernovae.}",
      journal = {\apj},
         year = 1984,
        month = feb,
       volume = {277},
        pages = {355-360},
          doi = {10.1086/161701},
       adsurl = {https://ui.adsabs.harvard.edu/abs/1984ApJ...277..355W}
}

@ARTICLE{Stroeer2007,
       author = {{Stroeer}, A. and {Heber}, U. and {Lisker}, T. and {Napiwotzki}, R. and {Dreizler}, S. and {Christlieb}, N. and {Reimers}, D.},
        title = "{Hot subdwarfs from the ESO supernova Ia progenitor survey. II. Atmospheric parameters of subdwarf O stars}",
      journal = {\aap},
         year = 2007,
        month = jan,
       volume = {462},
       number = {1},
        pages = {269-280},
          doi = {10.1051/0004-6361:20065564},
archivePrefix = {arXiv},
       eprint = {astro-ph/0609718},
 primaryClass = {astro-ph},
       adsurl = {https://ui.adsabs.harvard.edu/abs/2007A&A...462..269S}
}

@ARTICLE{Lindegren2021,
       author = {{Lindegren}, L. and {Bastian}, U. and {Biermann}, M. and {Bombrun}, A. and {de Torres}, A. and {Gerlach}, E. and {Geyer}, R. and {Hern{\'a}ndez}, J. and {Hilger}, T. and {Hobbs}, D. and {Klioner}, S.~A. and {Lammers}, U. and {McMillan}, P.~J. and {Ramos-Lerate}, M. and {Steidelm{\"u}ller}, H. and {Stephenson}, C.~A. and {van Leeuwen}, F.},
        title = "{Gaia Early Data Release 3. Parallax bias versus magnitude, colour, and position}",
      journal = {\aap},
         year = 2021,
        month = may,
       volume = {649},
          eid = {A4},
        pages = {A4},
          doi = {10.1051/0004-6361/202039653},
archivePrefix = {arXiv},
       eprint = {2012.01742},
 primaryClass = {astro-ph.IM},
       adsurl = {https://ui.adsabs.harvard.edu/abs/2021A&A...649A...4L}
}

@INPROCEEDINGS{Houck2000,
       author = {{Houck}, J.~C. and {Denicola}, L.~A.},
        title = "{ISIS: An Interactive Spectral Interpretation System for High Resolution X-Ray Spectroscopy}",
    booktitle = {Astronomical Data Analysis Software and Systems IX},
         year = 2000,
       editor = {{Manset}, Nadine and {Veillet}, Christian and {Crabtree}, Dennis},
       series = {Astronomical Society of the Pacific Conference Series},
       volume = {216},
        month = jan,
        pages = {591},
       adsurl = {https://ui.adsabs.harvard.edu/abs/2000ASPC..216..591H}
}

@ARTICLE{GaiaDR3,
       author = {{Gaia Collaboration} and {Vallenari}, A. and {Brown}, A.~G.~A. and {Prusti}, T. and {de Bruijne}, J.~H.~J. and {Arenou}, F. and {Babusiaux}, C. and {Biermann}, M. and {Creevey}, O.~L. and {Ducourant}, C. and {Evans}, D.~W. and {Eyer}, L. and {Guerra}, R. and {Hutton}, A. and {Jordi}, C. and {Klioner}, S.~A. and {Lammers}, U.~L. and {Lindegren}, L. and {Luri}, X. and {Mignard}, F. and {Panem}, C. and {Pourbaix}, D. and {Randich}, S. and {Sartoretti}, P. and {Soubiran}, C. and {Tanga}, P. and {Walton}, N.~A. and {Bailer-Jones}, C.~A.~L. and {Bastian}, U. and {Drimmel}, R. and {Jansen}, F. and {Katz}, D. and {Lattanzi}, M.~G. and {van Leeuwen}, F. and {Bakker}, J. and {Cacciari}, C. and {Casta{\~n}eda}, J. and {De Angeli}, F. and {Fabricius}, C. and {Fouesneau}, M. and {Fr{\'e}mat}, Y. and {Galluccio}, L. and {Guerrier}, A. and {Heiter}, U. and {Masana}, E. and {Messineo}, R. and {Mowlavi}, N. and {Nicolas}, C. and {Nienartowicz}, K. and {Pailler}, F. and {Panuzzo}, P. and {Riclet}, F. and {Roux}, W. and {Seabroke}, G.~M. and {Sordo}, R. and {Th{\'e}venin}, F. and {Gracia-Abril}, G. and {Portell}, J. and {Teyssier}, D. and {Altmann}, M. and {Andrae}, R. and {Audard}, M. and {Bellas-Velidis}, I. and {Benson}, K. and {Berthier}, J. and {Blomme}, R. and {Burgess}, P.~W. and {Busonero}, D. and {Busso}, G. and {C{\'a}novas}, H. and {Carry}, B. and {Cellino}, A. and {Cheek}, N. and {Clementini}, G. and {Damerdji}, Y. and {Davidson}, M. and {de Teodoro}, P. and {Nu{\~n}ez Campos}, M. and {Delchambre}, L. and {Dell'Oro}, A. and {Esquej}, P. and {Fern{\'a}ndez-Hern{\'a}ndez}, J. and {Fraile}, E. and {Garabato}, D. and {Garc{\'\i}a-Lario}, P. and {Gosset}, E. and {Haigron}, R. and {Halbwachs}, J. -L. and {Hambly}, N.~C. and {Harrison}, D.~L. and {Hern{\'a}ndez}, J. and {Hestroffer}, D. and {Hodgkin}, S.~T. and {Holl}, B. and {Jan{\ss}en}, K. and {Jevardat de Fombelle}, G. and {Jordan}, S. and {Krone-Martins}, A. and {Lanzafame}, A.~C. and {L{\"o}ffler}, W. and {Marchal}, O. and {Marrese}, P.~M. and {Moitinho}, A. and {Muinonen}, K. and {Osborne}, P. and {Pancino}, E. and {Pauwels}, T. and {Recio-Blanco}, A. and {Reyl{\'e}}, C. and {Riello}, M. and {Rimoldini}, L. and {Roegiers}, T. and {Rybizki}, J. and {Sarro}, L.~M. and {Siopis}, C. and {Smith}, M. and {Sozzetti}, A. and {Utrilla}, E. and {van Leeuwen}, M. and {Abbas}, U. and {{\'A}brah{\'a}m}, P. and {Abreu Aramburu}, A. and {Aerts}, C. and {Aguado}, J.~J. and {Ajaj}, M. and {Aldea-Montero}, F. and {Altavilla}, G. and {{\'A}lvarez}, M.~A. and {Alves}, J. and {Anders}, F. and {Anderson}, R.~I. and {Anglada Varela}, E. and {Antoja}, T. and {Baines}, D. and {Baker}, S.~G. and {Balaguer-N{\'u}{\~n}ez}, L. and {Balbinot}, E. and {Balog}, Z. and {Barache}, C. and {Barbato}, D. and {Barros}, M. and {Barstow}, M.~A. and {Bartolom{\'e}}, S. and {Bassilana}, J. -L. and {Bauchet}, N. and {Becciani}, U. and {Bellazzini}, M. and {Berihuete}, A. and {Bernet}, M. and {Bertone}, S. and {Bianchi}, L. and {Binnenfeld}, A. and {Blanco-Cuaresma}, S. and {Blazere}, A. and {Boch}, T. and {Bombrun}, A. and {Bossini}, D. and {Bouquillon}, S. and {Bragaglia}, A. and {Bramante}, L. and {Breedt}, E. and {Bressan}, A. and {Brouillet}, N. and {Brugaletta}, E. and {Bucciarelli}, B. and {Burlacu}, A. and {Butkevich}, A.~G. and {Buzzi}, R. and {Caffau}, E. and {Cancelliere}, R. and {Cantat-Gaudin}, T. and {Carballo}, R. and {Carlucci}, T. and {Carnerero}, M.~I. and {Carrasco}, J.~M. and {Casamiquela}, L. and {Castellani}, M. and {Castro-Ginard}, A. and {Chaoul}, L. and {Charlot}, P. and {Chemin}, L. and {Chiaramida}, V. and {Chiavassa}, A. and {Chornay}, N. and {Comoretto}, G. and {Contursi}, G. and {Cooper}, W.~J. and {Cornez}, T. and {Cowell}, S. and {Crifo}, F. and {Cropper}, M. and {Crosta}, M. and {Crowley}, C. and {Dafonte}, C. and {Dapergolas}, A. and {David}, M. and {David}, P. and {de Laverny}, P. and {De Luise}, F. and {De March}, R. and {De Ridder}, J. and {de Souza}, R. and {de Torres}, A. and {del Peloso}, E.~F. and {del Pozo}, E. and {Delbo}, M. and {Delgado}, A. and {Delisle}, J. -B. and {Demouchy}, C. and {Dharmawardena}, T.~E. and {Di Matteo}, P. and {Diakite}, S. and {Diener}, C. and {Distefano}, E. and {Dolding}, C. and {Edvardsson}, B. and {Enke}, H. and {Fabre}, C. and {Fabrizio}, M. and {Faigler}, S. and {Fedorets}, G. and {Fernique}, P. and {Fienga}, A. and {Figueras}, F. and {Fournier}, Y. and {Fouron}, C. and {Fragkoudi}, F. and {Gai}, M. and {Garcia-Gutierrez}, A. and {Garcia-Reinaldos}, M. and {Garc{\'\i}a-Torres}, M. and {Garofalo}, A. and {Gavel}, A. and {Gavras}, P. and {Gerlach}, E. and {Geyer}, R. and {Giacobbe}, P. and {Gilmore}, G. and {Girona}, S. and {Giuffrida}, G. and {Gomel}, R. and {Gomez}, A. and {Gonz{\'a}lez-N{\'u}{\~n}ez}, J. and {Gonz{\'a}lez-Santamar{\'\i}a}, I. and {Gonz{\'a}lez-Vidal}, J.~J. and {Granvik}, M. and {Guillout}, P. and {Guiraud}, J. and {Guti{\'e}rrez-S{\'a}nchez}, R. and {Guy}, L.~P. and {Hatzidimitriou}, D. and {Hauser}, M. and {Haywood}, M. and {Helmer}, A. and {Helmi}, A. and {Sarmiento}, M.~H. and {Hidalgo}, S.~L. and {Hilger}, T. and {H{\l}adczuk}, N. and {Hobbs}, D. and {Holland}, G. and {Huckle}, H.~E. and {Jardine}, K. and {Jasniewicz}, G. and {Jean-Antoine Piccolo}, A. and {Jim{\'e}nez-Arranz}, {\'O}. and {Jorissen}, A. and {Juaristi Campillo}, J. and {Julbe}, F. and {Karbevska}, L. and {Kervella}, P. and {Khanna}, S. and {Kontizas}, M. and {Kordopatis}, G. and {Korn}, A.~J. and {K{\'o}sp{\'a}l}, {\'A}. and {Kostrzewa-Rutkowska}, Z. and {Kruszy{\'n}ska}, K. and {Kun}, M. and {Laizeau}, P. and {Lambert}, S. and {Lanza}, A.~F. and {Lasne}, Y. and {Le Campion}, J. -F. and {Lebreton}, Y. and {Lebzelter}, T. and {Leccia}, S. and {Leclerc}, N. and {Lecoeur-Taibi}, I. and {Liao}, S. and {Licata}, E.~L. and {Lindstr{\o}m}, H.~E.~P. and {Lister}, T.~A. and {Livanou}, E. and {Lobel}, A. and {Lorca}, A. and {Loup}, C. and {Madrero Pardo}, P. and {Magdaleno Romeo}, A. and {Managau}, S. and {Mann}, R.~G. and {Manteiga}, M. and {Marchant}, J.~M. and {Marconi}, M. and {Marcos}, J. and {Marcos Santos}, M.~M.~S. and {Mar{\'\i}n Pina}, D. and {Marinoni}, S. and {Marocco}, F. and {Marshall}, D.~J. and {Martin Polo}, L. and {Mart{\'\i}n-Fleitas}, J.~M. and {Marton}, G. and {Mary}, N. and {Masip}, A. and {Massari}, D. and {Mastrobuono-Battisti}, A. and {Mazeh}, T. and {McMillan}, P.~J. and {Messina}, S. and {Michalik}, D. and {Millar}, N.~R. and {Mints}, A. and {Molina}, D. and {Molinaro}, R. and {Moln{\'a}r}, L. and {Monari}, G. and {Mongui{\'o}}, M. and {Montegriffo}, P. and {Montero}, A. and {Mor}, R. and {Mora}, A. and {Morbidelli}, R. and {Morel}, T. and {Morris}, D. and {Muraveva}, T. and {Murphy}, C.~P. and {Musella}, I. and {Nagy}, Z. and {Noval}, L. and {Oca{\~n}a}, F. and {Ogden}, A. and {Ordenovic}, C. and {Osinde}, J.~O. and {Pagani}, C. and {Pagano}, I. and {Palaversa}, L. and {Palicio}, P.~A. and {Pallas-Quintela}, L. and {Panahi}, A. and {Payne-Wardenaar}, S. and {Pe{\~n}alosa Esteller}, X. and {Penttil{\"a}}, A. and {Pichon}, B. and {Piersimoni}, A.~M. and {Pineau}, F. -X. and {Plachy}, E. and {Plum}, G. and {Poggio}, E. and {Pr{\v{s}}a}, A. and {Pulone}, L. and {Racero}, E. and {Ragaini}, S. and {Rainer}, M. and {Raiteri}, C.~M. and {Rambaux}, N. and {Ramos}, P. and {Ramos-Lerate}, M. and {Re Fiorentin}, P. and {Regibo}, S. and {Richards}, P.~J. and {Rios Diaz}, C. and {Ripepi}, V. and {Riva}, A. and {Rix}, H. -W. and {Rixon}, G. and {Robichon}, N. and {Robin}, A.~C. and {Robin}, C. and {Roelens}, M. and {Rogues}, H.~R.~O. and {Rohrbasser}, L. and {Romero-G{\'o}mez}, M. and {Rowell}, N. and {Royer}, F. and {Ruz Mieres}, D. and {Rybicki}, K.~A. and {Sadowski}, G. and {S{\'a}ez N{\'u}{\~n}ez}, A. and {Sagrist{\`a} Sell{\'e}s}, A. and {Sahlmann}, J. and {Salguero}, E. and {Samaras}, N. and {Sanchez Gimenez}, V. and {Sanna}, N. and {Santove{\~n}a}, R. and {Sarasso}, M. and {Schultheis}, M. and {Sciacca}, E. and {Segol}, M. and {Segovia}, J.~C. and {S{\'e}gransan}, D. and {Semeux}, D. and {Shahaf}, S. and {Siddiqui}, H.~I. and {Siebert}, A. and {Siltala}, L. and {Silvelo}, A. and {Slezak}, E. and {Slezak}, I. and {Smart}, R.~L. and {Snaith}, O.~N. and {Solano}, E. and {Solitro}, F. and {Souami}, D. and {Souchay}, J. and {Spagna}, A. and {Spina}, L. and {Spoto}, F. and {Steele}, I.~A. and {Steidelm{\"u}ller}, H. and {Stephenson}, C.~A. and {S{\"u}veges}, M. and {Surdej}, J. and {Szabados}, L. and {Szegedi-Elek}, E. and {Taris}, F. and {Taylor}, M.~B. and {Teixeira}, R. and {Tolomei}, L. and {Tonello}, N. and {Torra}, F. and {Torra}, J. and {Torralba Elipe}, G. and {Trabucchi}, M. and {Tsounis}, A.~T. and {Turon}, C. and {Ulla}, A. and {Unger}, N. and {Vaillant}, M.~V. and {van Dillen}, E. and {van Reeven}, W. and {Vanel}, O. and {Vecchiato}, A. and {Viala}, Y. and {Vicente}, D. and {Voutsinas}, S. and {Weiler}, M. and {Wevers}, T. and {Wyrzykowski}, {\L}. and {Yoldas}, A. and {Yvard}, P. and {Zhao}, H. and {Zorec}, J. and {Zucker}, S. and {Zwitter}, T.},
        title = "{Gaia Data Release 3. Summary of the content and survey properties}",
      journal = {\aap},
         year = 2023,
        month = jun,
       volume = {674},
          eid = {A1},
        pages = {A1},
          doi = {10.1051/0004-6361/202243940},
archivePrefix = {arXiv},
       eprint = {2208.00211},
 primaryClass = {astro-ph.GA},
       adsurl = {https://ui.adsabs.harvard.edu/abs/2023A&A...674A...1G}
}

@ARTICLE{ElBadry2021,
       author = {{El-Badry}, Kareem and {Rix}, Hans-Walter and {Heintz}, Tyler M.},
        title = "{A million binaries from Gaia eDR3: sample selection and validation of Gaia parallax uncertainties}",
      journal = {\mnras},
         year = 2021,
        month = sep,
       volume = {506},
       number = {2},
        pages = {2269-2295},
          doi = {10.1093/mnras/stab323},
archivePrefix = {arXiv},
       eprint = {2101.05282},
 primaryClass = {astro-ph.SR},
       adsurl = {https://ui.adsabs.harvard.edu/abs/2021MNRAS.506.2269E}
}

@ARTICLE{Zhang2012,
       author = {{Zhang}, Xianfei and {Jeffery}, C. Simon},
        title = "{AEvolutionary models for double helium white dwarf mergers and the formation of helium-rich hot subdwarfs}",
      journal = {\mnras},
         year = 2012,
        month = jan,
       volume = {419},
       number = {1},
        pages = {452-464},
          doi = {10.1111/j.1365-2966.2011.19711.x},
       adsurl = {https://ui.adsabs.harvard.edu/abs/2012MNRAS.419..452Z}
}

@ARTICLE{Zhang2012b,
       author = {{Zhang}, Xianfei and {Jeffery}, C. Simon},
        title = "{Can R Coronae Borealis stars form from the merger of two helium white dwarfs?}",
      journal = {\mnras},
         year = 2012,
        month = oct,
       volume = {426},
       number = {1},
        pages = {L81-L85},
          doi = {10.1111/j.1745-3933.2012.01330.x},
archivePrefix = {arXiv},
       eprint = {1208.3907},
 primaryClass = {astro-ph.SR},
       adsurl = {https://ui.adsabs.harvard.edu/abs/2012MNRAS.426L..81Z}
}

@PHDTHESIS{Dorsch2024,
       author = {{Dorsch}, Matti},
        title = "{Tracing the diversity of hot subdwarf evolution: Surface composition, magnetic fields, and populations}",
       school = {Friedrich Alexander University of Erlangen-Nuremberg, Germany},
         year = 2024,
        month = jan,
       adsurl = {https://ui.adsabs.harvard.edu/abs/2024PhDT........36D}
}

@dataset{GaiaBPRP2022,
       author = {{Gaia Collaboration}},
        title = "{VizieR Online Data Catalog: Gaia DR3 Part 1. Main source (Gaia Collaboration, 2022)}",
 howpublished = {VizieR On-line Data Catalog: I/355.  Originally published in: doi:10.1051/0004-63},
         year = 2022,
        month = may,
          eid = {I/355},
          doi = {10.26093/cds/vizier.1355},
       adsurl = {https://ui.adsabs.harvard.edu/abs/2022yCat.1355....0G}
}

@ARTICLE{VISTA1,
       author = {{Cioni}, M. -R.~L. and {Clementini}, G. and {Girardi}, L. and {Guandalini}, R. and {Gullieuszik}, M. and {Miszalski}, B. and {Moretti}, M. -I. and {Ripepi}, V. and {Rubele}, S. and {Bagheri}, G. and {Bekki}, K. and {Cross}, N. and {de Blok}, W.~J.~G. and {de Grijs}, R. and {Emerson}, J.~P. and {Evans}, C.~J. and {Gibson}, B. and {Gonzales-Solares}, E. and {Groenewegen}, M.~A.~T. and {Irwin}, M. and {Ivanov}, V.~D. and {Lewis}, J. and {Marconi}, M. and {Marquette}, J. -B. and {Mastropietro}, C. and {Moore}, B. and {Napiwotzki}, R. and {Naylor}, T. and {Oliveira}, J.~M. and {Read}, M. and {Sutorius}, E. and {van Loon}, J. Th. and {Wilkinson}, M.~I. and {Wood}, P.~R.},
        title = "{The VMC survey. I. Strategy and first data}",
      journal = {\aap},
         year = 2011,
        month = mar,
       volume = {527},
          eid = {A116},
        pages = {A116},
          doi = {10.1051/0004-6361/201016137},
archivePrefix = {arXiv},
       eprint = {1012.5193},
 primaryClass = {astro-ph.CO},
       adsurl = {https://ui.adsabs.harvard.edu/abs/2011A&A...527A.116C}
}

@ARTICLE{VISTA2,
       author = {{McMahon}, R.~G. and {Banerji}, M. and {Gonzalez}, E. and {Koposov}, S.~E. and {Bejar}, V.~J. and {Lodieu}, N. and {Rebolo}, R. and {VHS Collaboration}},
        title = "{First Scientific Results from the VISTA Hemisphere Survey (VHS)}",
      journal = {The Messenger},
         year = 2013,
        month = dec,
       volume = {154},
        pages = {35-37},
       adsurl = {https://ui.adsabs.harvard.edu/abs/2013Msngr.154...35M}
}

@ARTICLE{Skymapper2024,
       author = {{Onken}, Christopher A. and {Wolf}, Christian and {Bessell}, Michael S. and {Chang}, Seo-Won and {Luvaul}, Lance C. and {Tonry}, John L. and {White}, Marc C. and {Da Costa}, Gary S.},
        title = "{SkyMapper Southern Survey: Data release 4}",
      journal = {\pasa},
         year = 2024,
        month = oct,
       volume = {41},
          eid = {e061},
        pages = {e061},
          doi = {10.1017/pasa.2024.53},
archivePrefix = {arXiv},
       eprint = {2402.02015},
 primaryClass = {astro-ph.CO},
       adsurl = {https://ui.adsabs.harvard.edu/abs/2024PASA...41...61O}
}

@dataset{Denis2005,
       author = {{DENIS Consortium}},
        title = "{VizieR Online Data Catalog: The DENIS database (DENIS Consortium, 2005)}",
 howpublished = {VizieR On-line Data Catalog: B/denis.  Originally published in: 2005yCat.2263....0T},
         year = 2005,
        month = sep,
          eid = {B/denis},
       adsurl = {https://ui.adsabs.harvard.edu/abs/2005yCat....102002D}
}

@dataset{2Mass2003,
       author = {{Cutri}, R.~M. and {Skrutskie}, M.~F. and {van Dyk}, S. and {Beichman}, C.~A. and {Carpenter}, J.~M. and {Chester}, T. and {Cambresy}, L. and {Evans}, T. and {Fowler}, J. and {Gizis}, J. and {Howard}, E. and {Huchra}, J. and {Jarrett}, T. and {Kopan}, E.~L. and {Kirkpatrick}, J.~D. and {Light}, R.~M. and {Marsh}, K.~A. and {McCallon}, H. and {Schneider}, S. and {Stiening}, R. and {Sykes}, M. and {Weinberg}, M. and {Wheaton}, W.~A. and {Wheelock}, S. and {Zacarias}, N.},
        title = "{VizieR Online Data Catalog: 2MASS All-Sky Catalog of Point Sources (Cutri+ 2003)}",
 howpublished = {VizieR On-line Data Catalog: II/246.  Originally published in: University of Massachusetts and Infrared Processing and Analysis Center, (IPAC/California Institute of Technology) (2003)},
         year = 2003,
        month = jun,
          eid = {II/246},
       adsurl = {https://ui.adsabs.harvard.edu/abs/2003yCat.2246....0C}
}

@ARTICLE{WISE2,
       author = {{Schlafly}, Edward F. and {Meisner}, Aaron M. and {Green}, Gregory M.},
        title = "{The unWISE Catalog: Two Billion Infrared Sources from Five Years of WISE Imaging}",
      journal = {\apjs},
         year = 2019,
        month = feb,
       volume = {240},
       number = {2},
          eid = {30},
        pages = {30},
          doi = {10.3847/1538-4365/aafbea},
archivePrefix = {arXiv},
       eprint = {1901.03337},
 primaryClass = {astro-ph.IM},
       adsurl = {https://ui.adsabs.harvard.edu/abs/2019ApJS..240...30S}
}

@ARTICLE{Drout2023,
       author = {{Drout}, M.~R. and {G{\"o}tberg}, Y. and {Ludwig}, B.~A. and {Groh}, J.~H. and {de Mink}, S.~E. and {O'Grady}, A.~J.~G. and {Smith}, N.},
        title = "{An observed population of intermediate-mass helium stars that have been stripped in binaries}",
      journal = {Science},
         year = 2023,
        month = dec,
       volume = {382},
       number = {6676},
        pages = {1287-1291},
          doi = {10.1126/science.ade4970},
archivePrefix = {arXiv},
       eprint = {2307.00061},
 primaryClass = {astro-ph.SR},
       adsurl = {https://ui.adsabs.harvard.edu/abs/2023Sci...382.1287D}
}

@ARTICLE{Goetberg2023,
       author = {{G{\"o}tberg}, Y. and {Drout}, M.~R. and {Ji}, A.~P. and {Groh}, J.~H. and {Ludwig}, B.~A. and {Crowther}, P.~A. and {Smith}, N. and {de Koter}, A. and {de Mink}, S.~E.},
        title = "{Stellar Properties of Observed Stars Stripped in Binaries in the Magellanic Clouds}",
      journal = {\apj},
         year = 2023,
        month = dec,
       volume = {959},
       number = {2},
          eid = {125},
        pages = {125},
          doi = {10.3847/1538-4357/ace5a3},
archivePrefix = {arXiv},
       eprint = {2307.00074},
 primaryClass = {astro-ph.SR},
       adsurl = {https://ui.adsabs.harvard.edu/abs/2023ApJ...959..125G}
}

@ARTICLE{Latour2015,
       author = {{Latour}, M. and {Fontaine}, G. and {Green}, E.~M. and {Brassard}, P.},
        title = "{A non-LTE analysis of the hot subdwarf O star BD+28{\textdegree}4211. II. The optical spectrum}",
      journal = {\aap},
         year = 2015,
        month = jul,
       volume = {579},
          eid = {A39},
        pages = {A39},
          doi = {10.1051/0004-6361/201525999},
archivePrefix = {arXiv},
       eprint = {1504.06417},
 primaryClass = {astro-ph.SR},
       adsurl = {https://ui.adsabs.harvard.edu/abs/2015A&A...579A..39L}
}

@ARTICLE{Pakmor2024,
       author = {{Pakmor}, R{\"u}diger and {Pelisoli}, Ingrid and {Justham}, Stephen and {Rajamuthukumar}, Abinaya S. and {R{\"o}pke}, Friedrich K. and {Schneider}, Fabian R.~N. and {de Mink}, Selma E. and {Ohlmann}, Sebastian T. and {Podsiadlowski}, Philipp and {Mor{\'a}n-Fraile}, Javier and {Vetter}, Marco and {Andrassy}, Robert},
        title = "{Large-scale ordered magnetic fields generated in mergers of helium white dwarfs}",
      journal = {\aap},
         year = 2024,
        month = nov,
       volume = {691},
          eid = {A179},
        pages = {A179},
          doi = {10.1051/0004-6361/202451352},
archivePrefix = {arXiv},
       eprint = {2407.02566},
 primaryClass = {astro-ph.SR},
       adsurl = {https://ui.adsabs.harvard.edu/abs/2024A&A...691A.179P}
}

@ARTICLE{Dorsch2022,
       author = {{Dorsch}, M. and {Reindl}, N. and {Pelisoli}, I. and {Heber}, U. and {Geier}, S. and {Istrate}, A.~G. and {Justham}, S.},
        title = "{Discovery of a highly magnetic He-sdO star from a double-degenerate binary merger}",
      journal = {\aap},
         year = 2022,
        month = feb,
       volume = {658},
          eid = {L9},
        pages = {L9},
          doi = {10.1051/0004-6361/202142880},
archivePrefix = {arXiv},
       eprint = {2201.08146},
 primaryClass = {astro-ph.SR},
       adsurl = {https://ui.adsabs.harvard.edu/abs/2022A&A...658L...9D}
}

@ARTICLE{Dorsch2024b,
       author = {{Dorsch}, M. and {Jeffery}, C.~S. and {Philip Monai}, A. and {Tout}, C.~A. and {Snowdon}, E.~J. and {Monageng}, I. and {Scott}, L.~J.~A. and {Miszalski}, B. and {Woolf}, V.~M.},
        title = "{ZDiscovery of three magnetic helium-rich hot subdwarfs with SALT}",
      journal = {\aap},
         year = 2024,
        month = nov,
       volume = {691},
          eid = {A165},
        pages = {A165},
          doi = {10.1051/0004-6361/202451306},
archivePrefix = {arXiv},
       eprint = {2410.02737},
 primaryClass = {astro-ph.SR},
       adsurl = {https://ui.adsabs.harvard.edu/abs/2024A&A...691A.165D}
}

@ARTICLE{Luo2021,
       author = {{Luo}, Yangping and {N{\'e}meth}, P{\'e}ter and {Wang}, Kun and {Wang}, Xi and {Han}, Zhanwen},
        title = "{Hot Subdwarf Atmospheric Parameters, Kinematics, and Origins Based on 1587 Hot Subdwarf Stars Observed in Gaia DR2 and LAMOST DR7}",
      journal = {\apjs},
         year = 2021,
        month = oct,
       volume = {256},
       number = {2},
          eid = {28},
        pages = {28},
          doi = {10.3847/1538-4365/ac11f6},
archivePrefix = {arXiv},
       eprint = {2107.09302},
 primaryClass = {astro-ph.SR},
       adsurl = {https://ui.adsabs.harvard.edu/abs/2021ApJS..256...28L}
}

@ARTICLE{Reindl2014,
       author = {{Reindl}, N. and {Rauch}, T. and {Werner}, K. and {Kruk}, J.~W. and {Todt}, H.},
        title = "{On helium-dominated stellar evolution: the mysterious role of the O(He)-type stars}",
      journal = {\aap},
         year = 2014,
        month = jun,
       volume = {566},
          eid = {A116},
        pages = {A116},
          doi = {10.1051/0004-6361/201423498},
archivePrefix = {arXiv},
       eprint = {1405.1589},
 primaryClass = {astro-ph.SR},
       adsurl = {https://ui.adsabs.harvard.edu/abs/2014A&A...566A.116R}
}

@ARTICLE{Culpan2022,
       author = {{Culpan}, R. and {Geier}, S. and {Reindl}, N. and {Pelisoli}, I. and {Gentile Fusillo}, N. and {Vorontseva}, A.},
        title = "{The population of hot subdwarf stars studied with Gaia. IV. Catalogues of hot subluminous stars based on Gaia EDR3}",
      journal = {\aap},
         year = 2022,
        month = jun,
       volume = {662},
          eid = {A40},
        pages = {A40},
          doi = {10.1051/0004-6361/202243337},
archivePrefix = {arXiv},
       eprint = {2203.07938},
 primaryClass = {astro-ph.SR},
       adsurl = {https://ui.adsabs.harvard.edu/abs/2022A&A...662A..40C}
}

@ARTICLE{Paczynski1971,
       author = {{Paczy{\'n}ski}, B.},
        title = "{Evolution of Single Stars. IV. Helium Stars}",
      journal = {\actaa},
         year = 1971,
        month = jan,
       volume = {21},
        pages = {1},
       adsurl = {https://ui.adsabs.harvard.edu/abs/1971AcA....21....1P}
}

@ARTICLE{Nemeth2012,
       author = {{N{\'e}meth}, P{\'e}ter and {Kawka}, Ad{\'e}la and {Vennes}, St{\'e}phane},
        title = "{A selection of hot subluminous stars in the GALEX survey - II. Subdwarf atmospheric parameters}",
      journal = {\mnras},
         year = 2012,
        month = dec,
       volume = {427},
       number = {3},
        pages = {2180-2211},
          doi = {10.1111/j.1365-2966.2012.22009.x},
archivePrefix = {arXiv},
       eprint = {1211.0323},
 primaryClass = {astro-ph.SR},
       adsurl = {https://ui.adsabs.harvard.edu/abs/2012MNRAS.427.2180N}
}

@ARTICLE{Jeffery2011,
       author = {{Jeffery}, C.~S. and {Karakas}, A.~I. and {Saio}, H.},
        title = "{Double white dwarf mergers and elemental surface abundances in extreme helium and R Coronae Borealis stars}",
      journal = {\mnras},
         year = 2011,
        month = jul,
       volume = {414},
       number = {4},
        pages = {3599-3616},
          doi = {10.1111/j.1365-2966.2011.18667.x},
archivePrefix = {arXiv},
       eprint = {1103.1556},
 primaryClass = {astro-ph.SR},
       adsurl = {https://ui.adsabs.harvard.edu/abs/2011MNRAS.414.3599J}
}

@ARTICLE{Lanz1997,
       author = {{Lanz}, Thierry and {Hubeny}, Ivan and {Heap}, Sara R.},
        title = "{Non-LTE Line-blanketed Model Atmospheres of Hot Stars. III. Hot Subdwarfs: The sdO Star BD +75{\textdegree}325}",
      journal = {\apj},
         year = 1997,
        month = aug,
       volume = {485},
       number = {2},
        pages = {843-858},
          doi = {10.1086/304455},
       adsurl = {https://ui.adsabs.harvard.edu/abs/1997ApJ...485..843L}
}

@ARTICLE{Jeffery2026,
       author = {{Jeffery}, C.~S. and {Dorsch}, M. and {Philip Monai}, A. and {Snowdon}, E.~J. and {Monageng}, I. and {Miszalski}, B.},
        title = "{The SALT survey of helium-rich hot subdwarfs: final sample and classification}",
      journal = {\mnras},
         year = 2026,
        month = feb,
       volume = {546},
       number = {1},
          eid = {staf2202},
        pages = {staf2202},
          doi = {10.1093/mnras/staf2202},
archivePrefix = {arXiv},
       eprint = {2512.10529},
 primaryClass = {astro-ph.SR},
       adsurl = {https://ui.adsabs.harvard.edu/abs/2026MNRAS.546f2202J}
}

@ARTICLE{Latour2026,
       author = {{Latour}, M. and {Green}, E.~M. and {Dorsch}, M. and {Van Grootel}, V. and {Chayer}, P. and {Charpinet}, S. and {Heber}, U. and {Randall}, S.~K. and {Ma}, X.-Y.},
        title = "{The Arizona-Montr{\'e}al Spectroscopic Survey of hot subluminous stars}",
      journal = {\aap},
         year = 2026,
        month = jan,
       volume = {705},
          eid = {A248},
        pages = {A248},
          doi = {10.1051/0004-6361/202557314},
       adsurl = {https://ui.adsabs.harvard.edu/abs/2026A&A...705A.248L}
}

@ARTICLE{Rajamuthukumar2025,
       author = {{Rajamuthukumar}, Abinaya Swaruba and {Bauer}, Evan B. and {Justham}, Stephen and {Pakmor}, R{\"u}diger and {de Mink}, Selma E. and {Neunteufel}, Patrick},
        title = "{Evolution of binaries containing a hot subdwarf and a white dwarf to double white dwarfs, and double detonation supernovae with hypervelocity runaway stars}",
      journal = {\aap},
         year = 2025,
        month = dec,
       volume = {704},
          eid = {A82},
        pages = {A82},
          doi = {10.1051/0004-6361/202554452},
archivePrefix = {arXiv},
       eprint = {2411.08099},
 primaryClass = {astro-ph.SR},
       adsurl = {https://ui.adsabs.harvard.edu/abs/2025A&A...704A..82R}
}

@ARTICLE{Heber2024,
       author = {{Heber}, Ulrich},
        title = "{Hot Subdwarf Stars}",
      journal = {arXiv e-prints},
         year = 2024,
        month = oct,
          eid = {arXiv:2410.11663},
        pages = {arXiv:2410.11663},
          doi = {10.48550/arXiv.2410.11663},
archivePrefix = {arXiv},
       eprint = {2410.11663},
 primaryClass = {astro-ph.SR},
       adsurl = {https://ui.adsabs.harvard.edu/abs/2024arXiv241011663H}
}

@ARTICLE{Dawson2026,
       author = {{Dawson}, H. and {Dorsch}, M. and {Geier}, S. and {Munday}, J. and {Pritzkuleit}, M. and {Heber}, U. and {Mattig}, F. and {Benitez-Palacios}, D. and {Vu{\v{c}}kovi{\'c}}, M. and {Pelisoli}, I. and {Deshmukh}, K. and {Bhat}, A. and {Kufleitner}, L. and {Uzundag}, M. and {Schaffenroth}, V. and {Reindl}, N. and {Culpan}, R. and {Raddi}, R. and {Antunes Amaral}, L. and {Istrate}, A.~G. and {Justham}, S. and {{\O}stensen}, R.~H. and {Telting}, J.~H. and {Steinmetz}, T. and {Rodr{\'\i}guez-Segovia}, N. and {Fernandez-Schlosser}, P. and {Dur{\'a}n-Reyes}, A. and {Arancibia-Rojas}, E. and {Latour}, M. and {Jones}, G.~T. and {O'Brien}, M. and {Sahu}, S. and {Elms}, A.},
        title = "{A 500 pc volume-limited sample of hot subluminous stars: II. Atmospheric parameters, mass distribution, and kinematics}",
      journal = {\aap},
         year = 2026,
        month = feb,
       volume = {707},
          eid = {A6},
        pages = {A6},
          doi = {10.1051/0004-6361/202558123},
archivePrefix = {arXiv},
       eprint = {2601.02250},
 primaryClass = {astro-ph.SR},
       adsurl = {https://ui.adsabs.harvard.edu/abs/2026A&A...707A...6D}
}

@ARTICLE{Nepal2024,
       author = {{Nepal}, S. and {Chiappini}, C. and {Guiglion}, G. and {Steinmetz}, M. and {P{\'e}rez-Villegas}, A. and {Queiroz}, A.~B. and {Miglio}, A. and {Dohme}, P. and {Khalatyan}, A.},
        title = "{Insights from super-metal-rich stars: Is the Milky Way bar young?}",
      journal = {\aap},
         year = 2024,
        month = jan,
       volume = {681},
          eid = {L8},
        pages = {L8},
          doi = {10.1051/0004-6361/202348365},
archivePrefix = {arXiv},
       eprint = {2311.16993},
 primaryClass = {astro-ph.GA},
       adsurl = {https://ui.adsabs.harvard.edu/abs/2024A&A...681L...8N}
}

@ARTICLE{Blouin2021,
       author = {{Blouin}, Simon and {Daligault}, J{\'e}r{\^o}me and {Saumon}, Didier},
        title = "{$^{22}$Ne Phase Separation as a Solution to the Ultramassive White Dwarf Cooling Anomaly}",
      journal = {\apjl},
         year = 2021,
        month = apr,
       volume = {911},
       number = {1},
          eid = {L5},
        pages = {L5},
          doi = {10.3847/2041-8213/abf14b},
archivePrefix = {arXiv},
       eprint = {2103.12892},
 primaryClass = {astro-ph.SR},
       adsurl = {https://ui.adsabs.harvard.edu/abs/2021ApJ...911L...5B}
}

@ARTICLE{Bedard2024,
       author = {{B{\'e}dard}, Antoine and {Blouin}, Simon and {Cheng}, Sihao},
        title = "{Buoyant crystals halt the cooling of white dwarf stars}",
      journal = {\nat},
         year = 2024,
        month = mar,
       volume = {627},
       number = {8003},
        pages = {286-288},
          doi = {10.1038/s41586-024-07102-y},
archivePrefix = {arXiv},
       eprint = {2409.04419},
 primaryClass = {astro-ph.SR},
       adsurl = {https://ui.adsabs.harvard.edu/abs/2024Natur.627..286B}
}

@ARTICLE{GaiaDR2HRD2018,
       author = {{Gaia Collaboration} and {Babusiaux}, C. and {van Leeuwen}, F. and {Barstow}, M.~A. and {Jordi}, C. and {Vallenari}, A. and {Bossini}, D. and {Bressan}, A. and {Cantat-Gaudin}, T. and {van Leeuwen}, M. and {Brown}, A.~G.~A. and {Prusti}, T. and {de Bruijne}, J.~H.~J. and {Bailer-Jones}, C.~A.~L. and {Biermann}, M. and {Evans}, D.~W. and {Eyer}, L. and {Jansen}, F. and {Klioner}, S.~A. and {Lammers}, U. and {Lindegren}, L. and {Luri}, X. and {Mignard}, F. and {Panem}, C. and {Pourbaix}, D. and {Randich}, S. and {Sartoretti}, P. and {Siddiqui}, H.~I. and {Soubiran}, C. and {Walton}, N.~A. and {Arenou}, F. and {Bastian}, U. and {Cropper}, M. and {Drimmel}, R. and {Katz}, D. and {Lattanzi}, M.~G. and {Bakker}, J. and {Cacciari}, C. and {Casta{\~n}eda}, J. and {Chaoul}, L. and {Cheek}, N. and {De Angeli}, F. and {Fabricius}, C. and {Guerra}, R. and {Holl}, B. and {Masana}, E. and {Messineo}, R. and {Mowlavi}, N. and {Nienartowicz}, K. and {Panuzzo}, P. and {Portell}, J. and {Riello}, M. and {Seabroke}, G.~M. and {Tanga}, P. and {Th{\'e}venin}, F. and {Gracia-Abril}, G. and {Comoretto}, G. and {Garcia-Reinaldos}, M. and {Teyssier}, D. and {Altmann}, M. and {Andrae}, R. and {Audard}, M. and {Bellas-Velidis}, I. and {Benson}, K. and {Berthier}, J. and {Blomme}, R. and {Burgess}, P. and {Busso}, G. and {Carry}, B. and {Cellino}, A. and {Clementini}, G. and {Clotet}, M. and {Creevey}, O. and {Davidson}, M. and {De Ridder}, J. and {Delchambre}, L. and {Dell'Oro}, A. and {Ducourant}, C. and {Fern{\'a}ndez-Hern{\'a}ndez}, J. and {Fouesneau}, M. and {Fr{\'e}mat}, Y. and {Galluccio}, L. and {Garc{\'\i}a-Torres}, M. and {Gonz{\'a}lez-N{\'u}{\~n}ez}, J. and {Gonz{\'a}lez-Vidal}, J.~J. and {Gosset}, E. and {Guy}, L.~P. and {Halbwachs}, J.-L. and {Hambly}, N.~C. and {Harrison}, D.~L. and {Hern{\'a}ndez}, J. and {Hestroffer}, D. and {Hodgkin}, S.~T. and {Hutton}, A. and {Jasniewicz}, G. and {Jean-Antoine-Piccolo}, A. and {Jordan}, S. and {Korn}, A.~J. and {Krone-Martins}, A. and {Lanzafame}, A.~C. and {Lebzelter}, T. and {L{\"o}ffler}, W. and {Manteiga}, M. and {Marrese}, P.~M. and {Mart{\'\i}n-Fleitas}, J.~M. and {Moitinho}, A. and {Mora}, A. and {Muinonen}, K. and {Osinde}, J. and {Pancino}, E. and {Pauwels}, T. and {Petit}, J.-M. and {Recio-Blanco}, A. and {Richards}, P.~J. and {Rimoldini}, L. and {Robin}, A.~C. and {Sarro}, L.~M. and {Siopis}, C. and {Smith}, M. and {Sozzetti}, A. and {S{\"u}veges}, M. and {Torra}, J. and {van Reeven}, W. and {Abbas}, U. and {Abreu Aramburu}, A. and {Accart}, S. and {Aerts}, C. and {Altavilla}, G. and {{\'A}lvarez}, M.~A. and {Alvarez}, R. and {Alves}, J. and {Anderson}, R.~I. and {Andrei}, A.~H. and {Anglada Varela}, E. and {Antiche}, E. and {Antoja}, T. and {Arcay}, B. and {Astraatmadja}, T.~L. and {Bach}, N. and {Baker}, S.~G. and {Balaguer-N{\'u}{\~n}ez}, L. and {Balm}, P. and {Barache}, C. and {Barata}, C. and {Barbato}, D. and {Barblan}, F. and {Barklem}, P.~S. and {Barrado}, D. and {Barros}, M. and {Bartholom{\'e} Mu{\~n}oz}, L. and {Bassilana}, J.-L. and {Becciani}, U. and {Bellazzini}, M. and {Berihuete}, A. and {Bertone}, S. and {Bianchi}, L. and {Bienaym{\'e}}, O. and {Blanco-Cuaresma}, S. and {Boch}, T. and {Boeche}, C. and {Bombrun}, A. and {Borrachero}, R. and {Bouquillon}, S. and {Bourda}, G. and {Bragaglia}, A. and {Bramante}, L. and {Breddels}, M.~A. and {Brouillet}, N. and {Br{\"u}semeister}, T. and {Brugaletta}, E. and {Bucciarelli}, B. and {Burlacu}, A. and {Busonero}, D. and {Butkevich}, A.~G. and {Buzzi}, R. and {Caffau}, E. and {Cancelliere}, R. and {Cannizzaro}, G. and {Carballo}, R. and {Carlucci}, T. and {Carrasco}, J.~M. and {Casamiquela}, L. and {Castellani}, M. and {Castro-Ginard}, A. and {Charlot}, P. and {Chemin}, L. and {Chiavassa}, A. and {Cocozza}, G. and {Costigan}, G. and {Cowell}, S. and {Crifo}, F. and {Crosta}, M. and {Crowley}, C. and {Cuypers}, J. and {Dafonte}, C. and {Damerdji}, Y. and {Dapergolas}, A. and {David}, P. and {David}, M. and {de Laverny}, P.},
        title = "{Gaia Data Release 2. Observational Hertzsprung-Russell diagrams}",
      journal = {\aap},
         year = 2018,
        month = aug,
       volume = {616},
          eid = {A10},
        pages = {A10},
          doi = {10.1051/0004-6361/201832843},
archivePrefix = {arXiv},
       eprint = {1804.09378},
 primaryClass = {astro-ph.SR},
       adsurl = {https://ui.adsabs.harvard.edu/abs/2018A&A...616A..10G}
}

@ARTICLE{Cheng2019,
       author = {{Cheng}, Sihao and {Cummings}, Jeffrey D. and {M{\'e}nard}, Brice},
        title = "{A Cooling Anomaly of High-mass White Dwarfs}",
      journal = {\apj},
         year = 2019,
        month = dec,
       volume = {886},
       number = {2},
          eid = {100},
        pages = {100},
          doi = {10.3847/1538-4357/ab4989},
archivePrefix = {arXiv},
       eprint = {1905.12710},
 primaryClass = {astro-ph.SR},
       adsurl = {https://ui.adsabs.harvard.edu/abs/2019ApJ...886..100C}
}

@ARTICLE{Salaris2024,
       author = {{Salaris}, Maurizio and {Blouin}, Simon and {Cassisi}, Santi and {Bedin}, Luigi R.},
        title = "{$^{22}$Ne distillation and the cooling sequence of the old metal-rich open cluster NGC 6791}",
      journal = {\aap},
         year = 2024,
        month = jun,
       volume = {686},
          eid = {A153},
        pages = {A153},
          doi = {10.1051/0004-6361/202449209},
archivePrefix = {arXiv},
       eprint = {2403.02790},
 primaryClass = {astro-ph.SR},
       adsurl = {https://ui.adsabs.harvard.edu/abs/2024A&A...686A.153S}
}

@ARTICLE{Pritzkuleit2026,
       author = {{Pritzkuleit}, M. and {Dorsch}, M. and {Miller Bertolami}, M.~M. and {Geier}, S. and {Bradshaw}, C.~W. and {Dawson}, H.},
        title = "{Stars on the ascending helium giant branch: I. From white dwarf merger to helium giant: the evolutionary state of the rapidly rotating hot subdwarf HE 1518-0948}",
      journal = {\aap},
         year = 2026,
        month = jun,
       volume = {710},
          eid = {A99},
        pages = {A99},
          doi = {10.1051/0004-6361/202659703},
archivePrefix = {arXiv},
       eprint = {2604.11285},
 primaryClass = {astro-ph.SR},
       adsurl = {https://ui.adsabs.harvard.edu/abs/2026A&A...710A..99P}
}

@ARTICLE{Crawford2025,
       author = {{Crawford}, Courtney L. and {Li}, Yaguang and {Huber}, Daniel and {Yu}, Jie and {Bedding}, Timothy R. and {Martell}, Sarah L. and {Montet}, Benjamin T. and {Stello}, Dennis and {Isaacson}, Howard and {Howard}, Andrew W. and {Fulton}, Benjamin J. and {Zhang}, Jingwen and {Polanski}, Alex S. and {Weiss}, Lauren M.},
        title = "{The highest mass Kepler red giants ─ II. Spectroscopic parameters, the amplitude─activity relation, and unexpected halo orbits}",
      journal = {\mnras},
         year = 2025,
        month = oct,
       volume = {542},
       number = {4},
        pages = {3289-3301},
          doi = {10.1093/mnras/staf1421},
archivePrefix = {arXiv},
       eprint = {2508.12585},
 primaryClass = {astro-ph.SR},
       adsurl = {https://ui.adsabs.harvard.edu/abs/2025MNRAS.542.3289C}
}

@ARTICLE{Chen2025,
       author = {{Chen}, Kun and {Guo}, Yanjun and {Jiang}, Dengkai and {Han}, Zhanwen and {Chen}, Xuefei},
        title = "{The Binary Fraction of B-type Runaway Stars from LAMOST DR8}",
      journal = {\apj},
         year = 2025,
        month = aug,
       volume = {988},
       number = {2},
          eid = {228},
        pages = {228},
          doi = {10.3847/1538-4357/ade702},
archivePrefix = {arXiv},
       eprint = {2507.22569},
 primaryClass = {astro-ph.SR},
       adsurl = {https://ui.adsabs.harvard.edu/abs/2025ApJ...988..228C}
}

@ARTICLE{Gao2019,
       author = {{Gao}, Yan and {Li}, Jiao and {Jia}, Shi},
        title = "{High-velocity runaway binaries from supernovae in triple systems}",
      journal = {\mnras},
         year = 2019,
        month = aug,
       volume = {487},
       number = {3},
        pages = {3178-3182},
          doi = {10.1093/mnras/stz1525},
archivePrefix = {arXiv},
       eprint = {1905.12653},
 primaryClass = {astro-ph.SR},
       adsurl = {https://ui.adsabs.harvard.edu/abs/2019MNRAS.487.3178G}
}

@ARTICLE{Perets2012,
       author = {{Perets}, Hagai B. and {{\v{S}}ubr}, Ladislav},
        title = "{The Properties of Dynamically Ejected Runaway and Hyper-runaway Stars}",
      journal = {\apj},
         year = 2012,
        month = jun,
       volume = {751},
       number = {2},
          eid = {133},
        pages = {133},
          doi = {10.1088/0004-637X/751/2/133},
archivePrefix = {arXiv},
       eprint = {1202.2356},
 primaryClass = {astro-ph.SR},
       adsurl = {https://ui.adsabs.harvard.edu/abs/2012ApJ...751..133P}
}

@ARTICLE{Kroupa1998,
       author = {{Kroupa}, Pavel},
        title = "{On the binary properties and the spatial and kinematical distribution of young stars}",
      journal = {\mnras},
         year = 1998,
        month = jul,
       volume = {298},
       number = {1},
        pages = {231-242},
          doi = {10.1046/j.1365-8711.1998.01603.x},
archivePrefix = {arXiv},
       eprint = {astro-ph/9803157},
 primaryClass = {astro-ph},
       adsurl = {https://ui.adsabs.harvard.edu/abs/1998MNRAS.298..231K}
}

@ARTICLE{Dantas2023,
       author = {{Dantas}, M.~L.~L. and {Smiljanic}, R. and {Boesso}, R. and {Rocha-Pinto}, H.~J. and {Magrini}, L. and {Guiglion}, G. and {Tautvai{\v{s}}ien{\.{e}}}, G. and {Gilmore}, G. and {Randich}, S. and {Bensby}, T. and {Bragaglia}, A. and {Bergemann}, M. and {Carraro}, G. and {Jofr{\'e}}, P. and {Zaggia}, S.},
        title = "{The Gaia-ESO Survey: Old super-metal-rich visitors from the inner Galaxy}",
      journal = {\aap},
         year = 2023,
        month = jan,
       volume = {669},
          eid = {A96},
        pages = {A96},
          doi = {10.1051/0004-6361/202243667},
archivePrefix = {arXiv},
       eprint = {2210.08510},
 primaryClass = {astro-ph.GA},
       adsurl = {https://ui.adsabs.harvard.edu/abs/2023A&A...669A..96D}
}

@ARTICLE{MIST1,
       author = {{Dotter}, Aaron},
        title = "{MESA Isochrones and Stellar Tracks (MIST) 0: Methods for the Construction of Stellar Isochrones}",
      journal = {\apjs},
         year = 2016,
        month = jan,
       volume = {222},
       number = {1},
          eid = {8},
        pages = {8},
          doi = {10.3847/0067-0049/222/1/8},
archivePrefix = {arXiv},
       eprint = {1601.05144},
 primaryClass = {astro-ph.SR},
       adsurl = {https://ui.adsabs.harvard.edu/abs/2016ApJS..222....8D}
}

@ARTICLE{MIST2,
       author = {{Choi}, Jieun and {Dotter}, Aaron and {Conroy}, Charlie and {Cantiello}, Matteo and {Paxton}, Bill and {Johnson}, Benjamin D.},
        title = "{Mesa Isochrones and Stellar Tracks (MIST). I. Solar-scaled Models}",
      journal = {\apj},
         year = 2016,
        month = jun,
       volume = {823},
       number = {2},
          eid = {102},
        pages = {102},
          doi = {10.3847/0004-637X/823/2/102},
archivePrefix = {arXiv},
       eprint = {1604.08592},
 primaryClass = {astro-ph.SR},
       adsurl = {https://ui.adsabs.harvard.edu/abs/2016ApJ...823..102C}
}

@ARTICLE{Goetberg2017,
       author = {{G{\"o}tberg}, Y. and {de Mink}, S.~E. and {Groh}, J.~H.},
        title = "{Ionizing spectra of stars that lose their envelope through interaction with a binary companion: role of metallicity}",
      journal = {\aap},
         year = 2017,
        month = nov,
       volume = {608},
          eid = {A11},
        pages = {A11},
          doi = {10.1051/0004-6361/201730472},
archivePrefix = {arXiv},
       eprint = {1701.07439},
 primaryClass = {astro-ph.SR},
       adsurl = {https://ui.adsabs.harvard.edu/abs/2017A&A...608A..11G}
}

@ARTICLE{Krticka2024,
       author = {{Krti{\v{c}}ka}, J. and {Krti{\v{c}}kov{\'a}}, I. and {Jan{\'\i}k}, J. and {N{\'e}meth}, P. and {Kub{\'a}t}, J. and {Vu{\v{c}}kovi{\'c}}, M.},
        title = "{Hot subdwarf wind models with accurate abundances. II. Helium-dominated merger products CD-46 8926 and CD-51 11879}",
      journal = {\aap},
         year = 2024,
        month = mar,
       volume = {683},
          eid = {A80},
        pages = {A80},
          doi = {10.1051/0004-6361/202347978},
archivePrefix = {arXiv},
       eprint = {2401.04288},
 primaryClass = {astro-ph.SR},
       adsurl = {https://ui.adsabs.harvard.edu/abs/2024A&A...683A..80K}
}

@ARTICLE{MHorn2026,
       author = {{M{\"u}ller-Horn}, Johanna and {Ramachandran}, Varsha and {El-Badry}, Kareem and {Sander}, Andreas A.~C. and {Bodensteiner}, Julia and {Gies}, Douglas R. and {G{\"o}tberg}, Ylva and {Rivinius}, Thomas and {Shenar}, Tomer and {Sch{\"o}sser}, Elisa C. and {Wang}, Luqian and {Bieryla}, Allyson and {Buchhave}, Lars A. and {Latham}, David W.},
        title = "{Ultraviolet spectroscopy reveals a hot and luminous companion to the Be star+black hole candidate MWC 656}",
      journal = {\aap},
         year = 2026,
        month = apr,
       volume = {708},
          eid = {A187},
        pages = {A187},
          doi = {10.1051/0004-6361/202557960},
archivePrefix = {arXiv},
       eprint = {2601.14403},
 primaryClass = {astro-ph.SR},
       adsurl = {https://ui.adsabs.harvard.edu/abs/2026A&A...708A.187M}
}

@ARTICLE{Ramachandran2024,
       author = {{Ramachandran}, V. and {Sander}, A.~A.~C. and {Pauli}, D. and {Klencki}, J. and {Backs}, F. and {Tramper}, F. and {Bernini-Peron}, M. and {Crowther}, P. and {Hamann}, W.-R. and {Ignace}, R. and {Kuiper}, R. and {Oey}, M.~S. and {Oskinova}, L.~M. and {Shenar}, T. and {Todt}, H. and {Vink}, J.~S. and {Wang}, L. and {Wofford}, A. and {the XShootU Collaboration}},
        title = "{X-Shooting ULLYSES: Massive stars at low metallicity: VIII. Stellar and wind parameters of newly revealed stripped stars in Be binaries}",
      journal = {\aap},
         year = 2024,
        month = dec,
       volume = {692},
          eid = {A90},
        pages = {A90},
          doi = {10.1051/0004-6361/202449665},
archivePrefix = {arXiv},
       eprint = {2406.17678},
 primaryClass = {astro-ph.SR},
       adsurl = {https://ui.adsabs.harvard.edu/abs/2024A&A...692A..90R}
}

@ARTICLE{Ramachandran2023,
       author = {{Ramachandran}, V. and {Klencki}, J. and {Sander}, A.~A.~C. and {Pauli}, D. and {Shenar}, T. and {Oskinova}, L.~M. and {Hamann}, W.-R.},
        title = "{A partially stripped massive star in a Be binary at low metallicity. A missing link towards Be X-ray binaries and double neutron star mergers}",
      journal = {\aap},
         year = 2023,
        month = jun,
       volume = {674},
          eid = {L12},
        pages = {L12},
          doi = {10.1051/0004-6361/202346818},
archivePrefix = {arXiv},
       eprint = {2305.14262},
 primaryClass = {astro-ph.SR},
       adsurl = {https://ui.adsabs.harvard.edu/abs/2023A&A...674L..12R}
}

@ARTICLE{Chojnowski2018,
       author = {{Chojnowski}, S. Drew and {Labadie-Bartz}, Jonathan and {Rivinius}, Thomas and {Gies}, Douglas and {Panoglou}, Despina and {Borges Fernandes}, Marcelo and {Wisniewski}, John P. and {Whelan}, David G. and {Mennickent}, Ronald E. and {McMillan}, Russet and {Dembicky}, Jack M. and {Gray}, Candace and {Rudyk}, Ted and {Stringfellow}, Guy S. and {Lester}, Kathryn and {Hasselquist}, Sten and {Zharikov}, Sergey and {Levenhagen}, Ronaldo and {Souza}, Tiago and {Leister}, Nelson and {Stassun}, Keivan and {Siverd}, Robert J. and {Majewski}, Steven R.},
        title = "{The Remarkable Be+sdOB Binary HD 55606. I. Orbital and Stellar Parameters}",
      journal = {\apj},
         year = 2018,
        month = sep,
       volume = {865},
       number = {1},
          eid = {76},
        pages = {76},
          doi = {10.3847/1538-4357/aad964},
archivePrefix = {arXiv},
       eprint = {1806.06843},
 primaryClass = {astro-ph.SR},
       adsurl = {https://ui.adsabs.harvard.edu/abs/2018ApJ...865...76C}
}

@ARTICLE{Wang2021,
       author = {{Wang}, Luqian and {Gies}, Douglas R. and {Peters}, Geraldine J. and {G{\"o}tberg}, Ylva and {Chojnowski}, S. Drew and {Lester}, Kathryn V. and {Howell}, Steve B.},
        title = "{The Detection and Characterization of Be+sdO Binaries from HST/STIS FUV Spectroscopy}",
      journal = {\aj},
         year = 2021,
        month = may,
       volume = {161},
       number = {5},
          eid = {248},
        pages = {248},
          doi = {10.3847/1538-3881/abf144},
archivePrefix = {arXiv},
       eprint = {2103.13642},
 primaryClass = {astro-ph.SR},
       adsurl = {https://ui.adsabs.harvard.edu/abs/2021AJ....161..248W}
}

@ARTICLE{ArancibiaRojas2024,
       author = {{Arancibia-Rojas}, Eduardo and {Zorotovic}, Monica and {Vu{\v{c}}kovi{\'c}}, Maja and {Bobrick}, Alexey and {Vos}, Joris and {Piraino-Cerda}, Franco},
        title = "{The mass range of hot subdwarf B stars from MESA simulations}",
      journal = {\mnras},
         year = 2024,
        month = feb,
       volume = {527},
       number = {4},
        pages = {11184-11197},
          doi = {10.1093/mnras/stad3891},
archivePrefix = {arXiv},
       eprint = {2312.09920},
 primaryClass = {astro-ph.SR},
       adsurl = {https://ui.adsabs.harvard.edu/abs/2024MNRAS.52711184A}
}

\begin{appendix}

\section{Light curve}\label{sect_lcurve}

\begin{figure}
\centering
\includegraphics[width=0.8\linewidth]{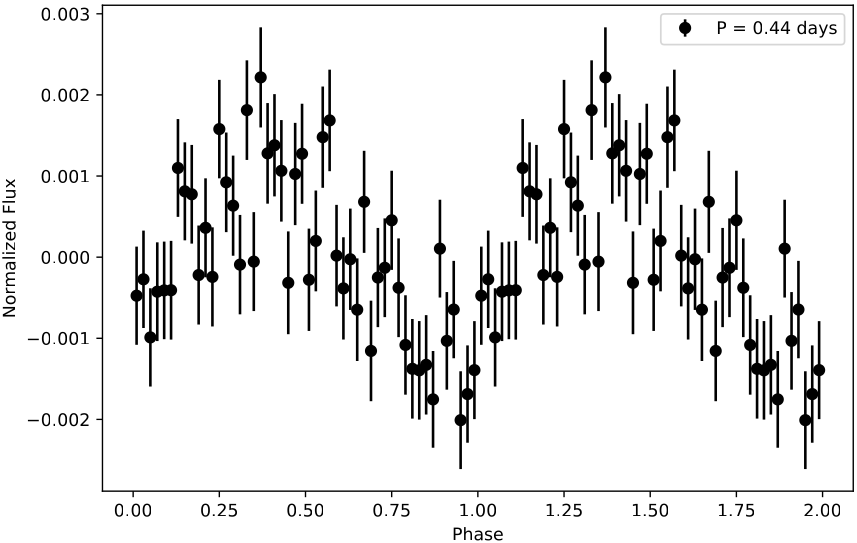}
\caption{TESS Light curve of G5369.}
\label{lightcurve}
\FloatBarrier
\end{figure}

To identify a close stellar companion or pulsation we checked for light curve data obtained by the Transiting Exoplanet Survey Satellite (TESS). Light curves were available for G5369 with a cadence of 200~s from the MIT Quick Look Pipeline (QLP; \citealt{Huang2020}) using data obtained in sectors 64, 90, and 100. We computed a Lomb-Scargle periodogram on the individual sectors using the
flux determined from the smallest extraction aperture in the QLP. We consider a signal to be reliable when it lies above a false alarm probability of 0.1\% in the periodogram. Only in sector 100 did we detect a significant peak with a period of 10.64 hours. The combined phase-folded light curve is shown in Fig.~\ref{lightcurve}. However, two bright sources are located close to G5369 at angular separations of $21\arcsec$ (Gaia DR3 5369742905756076800) and $24\arcsec$ (Gaia DR3 5369742871396330880), with $G$ magnitudes of 13.64~mag and 13.23~mag, respectively. Given the TESS pixel scale of approximately $21\arcsec$, we also examined the light curves of these two sources. We found that the variability is present in the light curves of all three sources but is strongest for Gaia DR3 5369742905756076800, indicating that the signal most likely originates from this source.

Assuming that this signal is caused by a spot on the surface of G5369 it allows us to determine the rotational velocity independent of the inclination. The spot could result from an undetected magnetic field generated during the merger. \citet{Pakmor2024} predict the formation of magnetic fields in mergers of two equal-mass WDs, and magnetic fields have been identified in at least four intermediate He-rich sdOs \citep{Dorsch2022, Dorsch2024b}. Using the radius derived from the SED fit, we obtain a rotational velocity of $v_\mathrm{rot}=82^{+9}_{-8}~\mathrm{km~s}^{-1}$, which is very similar to that measured for HE~1518. However, this would require a relatively low inclination of $i=17\pm4^\circ$. Moreover, stars with magnetic fields are not expected to retain high rotational velocities because of magnetic braking.

\section{Additional tables and figures}

\begin{figure}[!h]
\centering
\includegraphics[width=0.8\columnwidth]{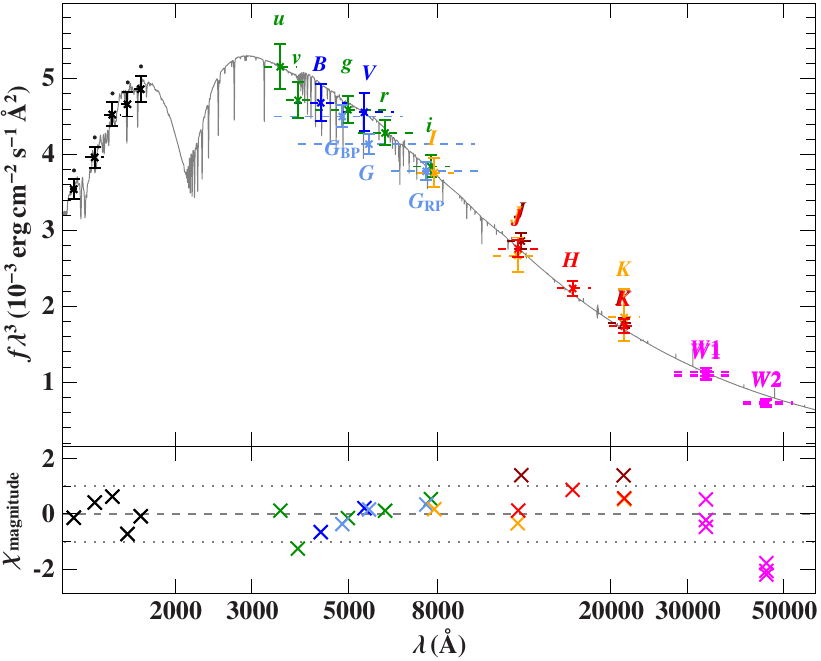}
%\hspace{5pt}
\caption{SED fitting performed using photometry from the following surveys covering the UV to infrared range: the HST/STIS spectrum divided into boxes (black), \textit{Gaia} $G$/$G_{BP}$/$G_{RP}$ photometry \citep{GaiaBPRP2022} (light blue), Johnson \citep{Henden2016} (blue), Skymapper \citep{Skymapper2024} (green), VISTA \citep{VISTA1, VISTA2} (maroon), DENIS \citep{Denis2005} (orange), 2MASS \citep{2Mass2003} (red), and WISE \citep{WISE2} (pink).}
\label{SED}
\FloatBarrier
\end{figure}

\begin{figure}
\centering
\includegraphics[width=0.8\linewidth]{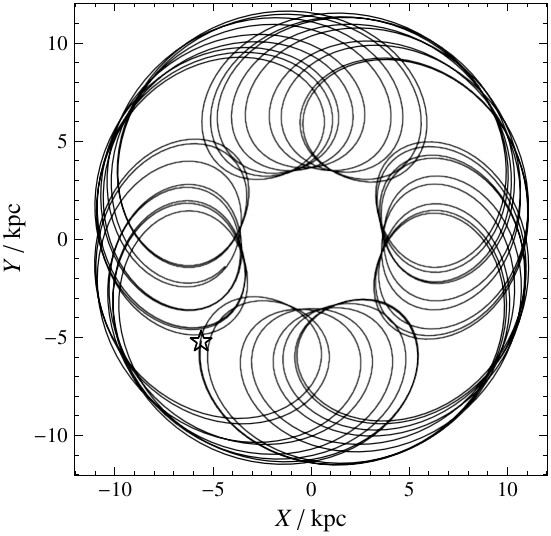}
\caption{Projection of the stellar orbit shown in Fig.~\ref{rhoz} onto the $X$--$Y$ plane in the co-rotating frame. The open star indicates the present-day position.}
\label{kinematics_corot}
\end{figure}

\begin{table}
\caption{Weak uncontaminated lines of Mg, Al, P, Cr, Mn, and Zn that were used to determine their abundances. }
\label{elements}      
\centering                                      
\begin{tabular}{l l}          
\toprule\toprule                     
Ion & Wavelength [\AA] \\ \midrule
Mg \textsc{iii} & 1572.71, 1687.08, 1697.27 \\
Al \textsc{iii}  &  1605.77, 1611.87 \\
P \textsc{v}    & 1610.50 \\
Cr \textsc{v}  &  1193.95,  1481.66, 1482.76, 1489.71, \\
Cr \textsc{v}  &   1497.97, 1519.03, 1579.69 \\
Mn \textsc{v}  &  1436.93, 1437.39, 1443.31  \\
Mn \textsc{v}  &  1502.33, 1503.58, 1504.34   \\
Zn \textsc{iv}  & 1174.32, 1176.87, 1176.98, 1177.04 \\
\bottomrule                                            
\end{tabular}
\FloatBarrier
\end{table}

\begin{figure*}
    \centering
    \includegraphics[page=1, width=0.94\linewidth]{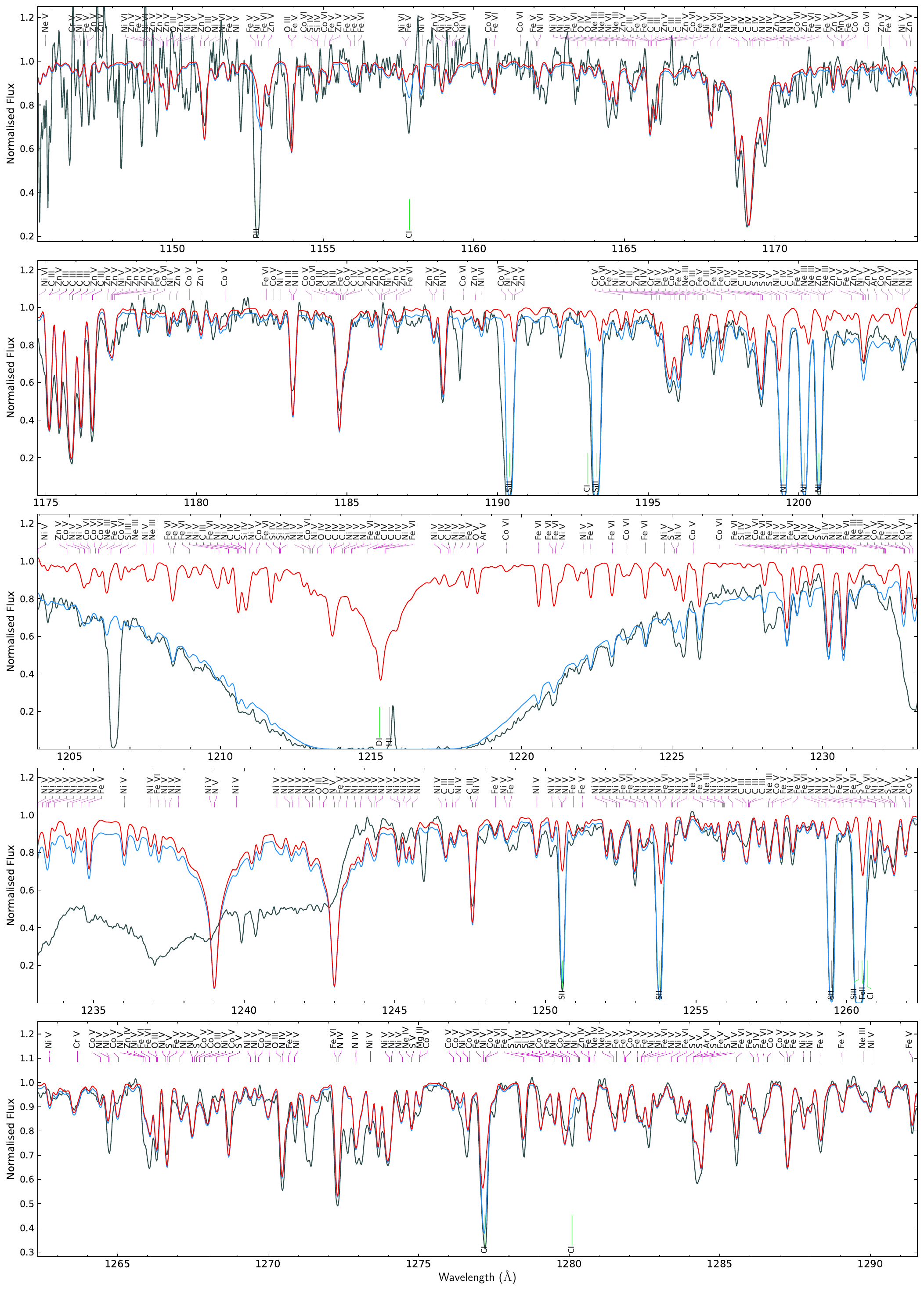}
    \caption{Complete STIS/E140M spectrum, shown at $R=20\,000$ for clarity. The colour coding is the same as in Fig.~\ref{Specfit}.}
    \label{fullspec}
\end{figure*}

\begin{figure*}
    \centering
    \includegraphics[page=2, width=0.94\linewidth]{GAIA_EDR3_5369742867090186624_STIS_HK.pdf}
     \addtocounter{figure}{-1}
    \caption{continued.}
\end{figure*}

\begin{figure*}
    \centering
    \includegraphics[page=3, width=0.94\linewidth]{GAIA_EDR3_5369742867090186624_STIS_HK.pdf}
     \addtocounter{figure}{-1}
    \caption{continued.}
\end{figure*}

\begin{figure*}
    \centering
    \includegraphics[page=4, width=0.94\linewidth]{GAIA_EDR3_5369742867090186624_STIS_HK.pdf}
     \addtocounter{figure}{-1}
    \caption{continued.}
\end{figure*}

\end{appendix}

\end{document}